\documentstyle[12pt,aasms4]{article}

\newcommand{\kms}{km~s$^{-1}$}
\newcommand\msun{$M_{\odot}$}
\newcommand\rsun{$R_{\odot}$}

\lefthead{Carney et al.}
\righthead{Halo Binary Deficiency}

\begin{document}

\title{A Survey of Proper Motion Stars. XVII. \\
 A Deficiency of Binary Stars on Retrograde Galactic Orbits
and the Possibility that $\omega$ Centauri is Related
to the Effect.}

\author{Bruce W. Carney}
\affil{Department of Physics \& Astronomy, University of 
North Carolina, Chapel Hill, NC 27599-3255; bruce@physics.unc.edu}

\author{Luis A. Aguilar}
\affil{Observatorio Astronomico Na\c{c}ional, 
Apdo.\ Postal 877, Ensenada, BC 22800, Mexico; 
aguilar@astrosen.unam.mx}

\author{David W. Latham}
\affil{Center for Astrophysics, 60 Garden Street, Cambridge, 
MA 02138; dlatham@cfa.harvard.edu}

\and 

\author{John B. Laird}
\affil{Department of Physics \& Astronomy, Bowling Green
State Univeristy, Bowling Green OH 43403; laird@tycho.bgsu.edu}

\begin{abstract}

We compare the frequency of field binary stars as a function of Galactic
velocity vectors, and find a deficiency of such stars on strongly
retrograde orbits. Metal-poor stars moving on
prograde Galactic orbits have a binary frequency of $28 \pm 3$\%,
while the retrograde stars' binary frequency is only $10 \pm 2$\%
for V $\leq -300$ \kms. No such binary deficiencies are seen for
the U or W velocities, nor [Fe/H]. Some mechanism exists that
either disrupts binary systems or preferentially
adds single stars moving primarily on retrograde orbits.

Theoretical analyses and critical evaluations of
our observational data appear to rule out preferential disruption
of pre-existing binary stars
due to such causes as
tidal interactions with massive gravitional perturbers, including
giant molecular clouds, black holes, or the Galactic center.

Dynamically-evolved stellar ensembles, such as globular clusters,
provide a possible source of single stars. Three lines of
evidence rule out this explanation. First, there is no mechanism to
significantly enhance dissolution of clusters moving on retrograde orbits.
Second, a study of globular clusters moving
on prograde and retrograde orbits, and with perigalacticon distances
such that they are unlikely to be affected strongly by central tidal
effects, shows that
clusters moving on prograde Galactic orbits may be more evolved
dynamically than clusters moving on retrograde orbits.
Finally, we have undertaken a comprehensive search for star
streams that might be disernible. Monte Carlo modelling suggests
that our sample may include one moving group, but it contains
only five stars. While the Galactic orbit of this group passes
near the Galactic center, it is not moving on a retrograde Galactic
orbit, and falls short by a factor of at least twenty in supplying
the necessary number of single stars.

There is one intriguing possibility to explain our results. A dissolved
dwarf galaxy may have too large a velocity spread to be easily detected
in our sample using our technique. But dwarf galaxies appear to
often show element-to-iron vs.\ [Fe/H] abundance patterns that are
not similar to the bulk of the stellar field and cluster halo stars.
We explore the $s$-process elements Y and Ba. Eight
stars in our sample have such elemental abundances
already measured and also lie in the critical
domain with  $-1.6 \leq$\ [Fe/H] $\leq\ -1.0$
and V $\leq\ -300$ \kms. The
admittedly small samples appears to show a bimodal distribution in
[Y/Fe], [Ba/Fe], and [$\alpha$/Fe], where ``$\alpha$"
represents an average abundance of Mg, Si, Ca, and Ti. 
This behavior is reminiscent of the difference
in the abundances found between the globular cluster $\omega$~Centauri
and other globular clusters. It is also intriguing that
the stars most similar to $\omega$~Cen in their chemical
abundances show a relatively coherent set of kinematic
properties, with a modest velocity dispersion. The stars less
like $\omega$~Cen define a dynamically hot population.
The binary frequency of the
stars in $\omega$~Centauri does not appear to be
enhanced, but detailed modelling of the radial velocity data
remains to be done.

\end{abstract}



\section{INTRODUCTION}

Binary and multiple stellar systems are common among stars
in the solar neighborhood (see Duquennoy \& Mayor 1991 and
references therein). While some earlier radial velocity
studies had suggested that the Galaxy's metal-poor, high-velocity
halo population was relatively deficient in binaries
(Abt \& Levy 1969; Crampton \& Hartwick 1972; Abt \& Willmarth 1987),
small sample sizes prevented any firm conclusions. Identification
of halo population binaries began to increase in the 1980s, with
orbital solutions for HD~111980 and HD~149414 (Mayor \& Turon 1982),
BD$-3$~2525 (Greenstein \& Saha 1986), BD+13~3683 (Jasniewicz \&
Mayor 1986), and CD$-48$~1741 (Lindgren, Ardeberg, \& Zuiderwijk 1987).

A much larger number of binary star orbits for halo stars,
as well as for stars belonging to the thick disk and thin disk populations,
are appearing as a consequence of long-term spectroscopic
monitoring of the stars in the ``Carney-Latham" survey
(hereafter CLLA program; 
Carney \& Latham 1987; Laird, Carney, \& Latham 1988a; Carney et al.\ 1994). 
Orbital solutions
for around two hundred stars have been published (Latham
et al.\ 1988, 1992, 2002; Goldberg et al.\ 2002;
Carney et al.\ 2001), with periods ranging from 1.9 to 7500 days.
For a combined system mass of 1.2 \msun, this corresponds
to major axis separations of about 0.03~AU (7~\rsun) to about 8~AU. 
Some even shorter period systems' orbital solutions have yet to
be reported (as short as 0.5 day) and a number of spectroscopic binaries
with longer periods but incomplete orbital solutions have also
been observed. Our goal has been to obtain at least 10
velocities over a span of at least 10 years for every
star, which we have reached for almost all the single stars.
(A few stars have only 8 or 9 observations.)
Binary stars typically have twenty or more
observations, but often over a shorter span since observing
is ended or much reduced in priority when the orbital solution
is considered solved. 

In Figure~7 of Latham et al.\ (2002), we showed how we roughly
divided the binary and constant velocity stars into disk and
halo components. It was our subjective impression that the
fraction of binary stars on retrograde Galactic orbits (V $<$ $-220$ \kms)
was lower than expected. This paper explores that issue in more
detail, and what it may reveal about the formation and evolution
of the Galactic halo.

We have also included here results from preliminary
spectroscopic binary orbital solutions
from a second study of metal-poor stars, also selected on
the basis of high proper motions.
In 1992 we began radial velocity monitoring of 472 metal-poor
stars selected from the sample of Ryan (1989; see also Ryan \& Norris 1991a).
Stars were selected if their estimated [Fe/H] values were $-0.4$ or lower
and the declinations were north of $-25$ degrees.
These data were obtained and analyzed in the same manner as reported by Carney
et al.\ (1994): Metallicities were obtained from $\chi^{2}$ comparisons
with a grid of synthetic spectra using a stellar effective temperature
determined via photometry (see Carney et al.\ 1987; Laird et al.\ 1988a). 
Details of the second study will be presented later. 

At the
time of writing, we had obtained 28,564 velocities of 1,464 stars
in the CLLA program, with a maximal span in the radial velocity coverage of
almost 20 years. 
For the Ryan sample, we had obtained 3,812 velocities
of 472 stars. Our goal for this second sample is much less ambitious since
we have been interested primarily in measuring the stars' metallicities,
which requires only a few spectra (at most), and simply detecting
shorter period binaries for follow-up observations. One goal,
for example, was to refine the ``transition period" that
separates spectroscopic binary systems with
circular orbits from those with eccentric orbits,
which occurs at an orbital period of around twenty days. We also
were searching for additional metal-poor double-lined systems that might
show eclipses.
While a few stars
have radial velocity coverage of up to ten years, the average
is much less, about two years, and in some cases the coverage is
only a month or two. The Ryan sample is thus not yet a proper
tool for detailed studies of binary statistics, but because the
stars were selected in a very different manner than those in the
CLLA survey, the Ryan sample provides a good check on the results
from the CLLA survey.

Within both surveys,
all stars recognized as velocity variables (with final
or preliminary orbital solutions, and those with secularly-changing
velocities indicative of orbital motion) are
identified here as ``binaries". To obtain a maximal
identification of binaries for this paper, we also included visual
binary stars and common proper motion pairs. To avoid
overcounting, all of the latter, even when one or both of the
members of the pair were themselves spectroscopic binaries,
were counted as just one binary. Therefore, the G72-58 and G72-59
common proper motion pair, each of which 
has a published single-lined spectroscopic binary orbital
solution, was counted as one binary at the relevant metallicity.
In Figures~1a,b we show the distribution of Galactic V velocities 
of binary and single stars
as a function of metallicity, [Fe/H], for [Fe/H] $\leq$\ $0.0$.
(Please note that we have generally reported metallicities as
``[m/H]" to indicate a mean metallicity rather than the abundance
of a specific element. However, the calibrations have been done
using [Fe/H], and to minimize confusion in comparisons with other
studies, we refer here to all of our [m/H] results as [Fe/H].)
Blue stragglers, subgiants, and all stars with significant
uncertainties which compromised the determination of either
metallicity or kinematics were rejected, leaving 756 single
stars and 238 binaries in the CLLA
program, and 349 single
stars and 63 binaries in the Ryan program. The combined
sample of 1406 stars has a binary fraction, $f$, of $21.4 \pm 1.2$\%,
where $f$ is defined as the number of binary stars divided by
the total sample size. NOTE: we do not claim here that this is
the ``true binary fraction" of the high proper motion stars
in our sample. Even with long-term radial velocity monitoring,
very long period systems will be overlooked, as will systems
with unfavorable orbital inclinations. No doubt
common proper motion pairs remain to be discovered as well.
Insofar as $f$ is determined without kinematic or
metallicity bias, we can use $f$
to explore variations in binary fractions as functions of
those parameters.

The calculation of stellar kinematics relies on relations
between reddening-corrected color indices, metallicity, and
absolute magnitudes, calibrated using stars in our surveys with
good parallax data from the Hipparcos satellite. Details will
be presented later, but we may estimate the approximate levels
of precision in three ways. First, we have 16 common proper motion
pairs of stars, for which we have determined the [Fe/H] values and
the U, V, and W velocities independently. The average value
of the velocity difference for the 48 separate
velocity component comparisons is 8 \kms,
with only 12 having differences exceeding 10 \kms, and only 5 with differences
greater than 20 \kms. Since identical proper motion values were employed,
errors in the proper motions themselves must contribute additional
uncertainties to our derived kinematics for individual
stars. Assuming that the total
proper motion error is less than 10\%, and that it is apportioned equally
among the three velocities, then to the 8 \kms\ uncertainty we should add to that
roughly 10 \kms\ for stars with individual velocities typical
of our program stars. Thus, independent of more ``global" and systematic
effects, including our derivations of [Fe/H] and $M_{V}$, individual UVW
velocities should be accurate to perhaps 15 \kms. A more conservative
choice would be 20 \kms.

\begin{figure}
\epsscale{1}
\figurenum{1a}
\plotone{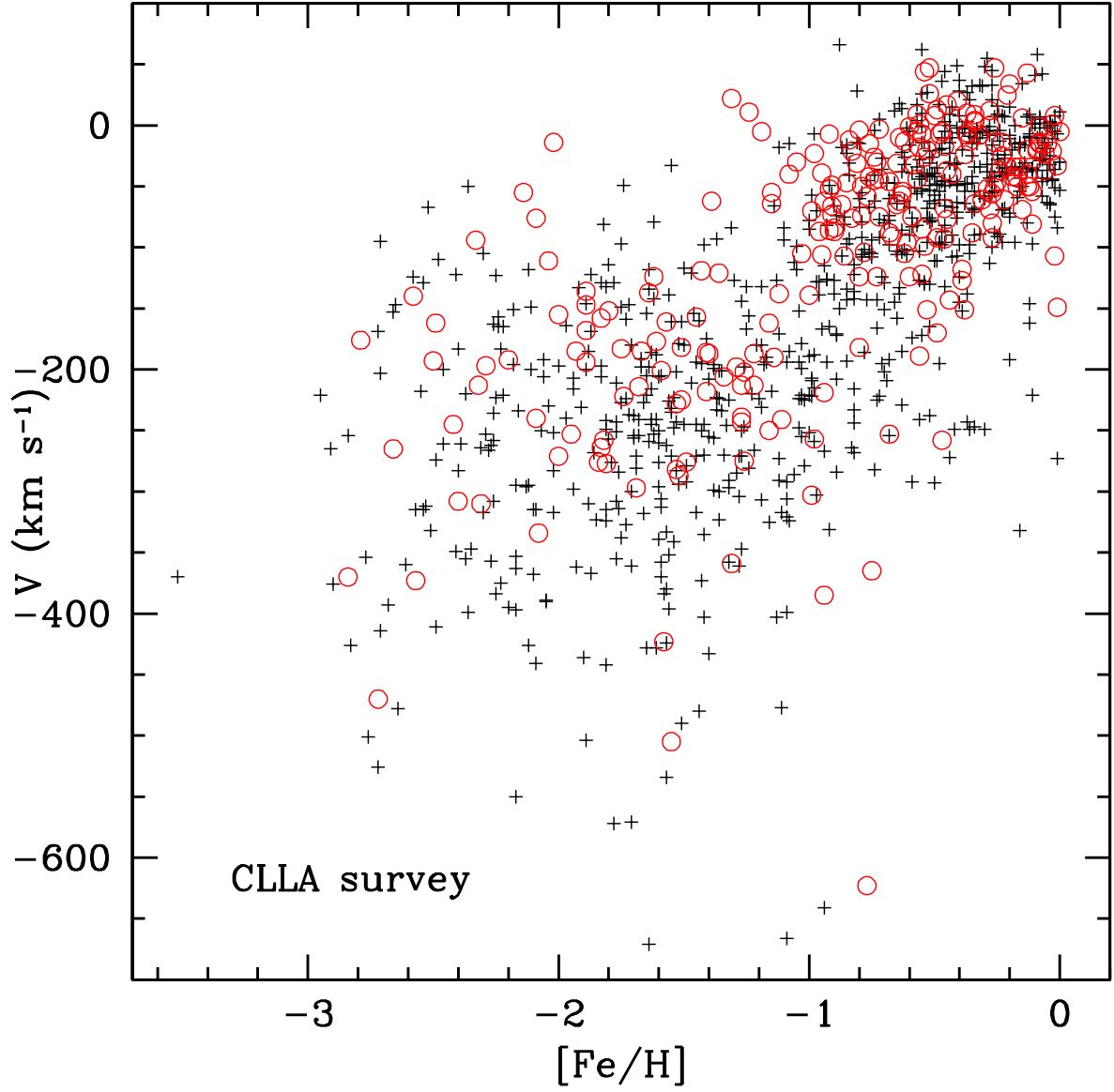}
\caption{(a) The distribution of 
single (+) and binary stars (o) in
the CLLA survey as a function of V velocity and metallicity, [Fe/H].
\label{fig1a}}
\end{figure}

\placefigure{fig1a}

\begin{figure}
\epsscale{1}
\figurenum{1b}
\plotone{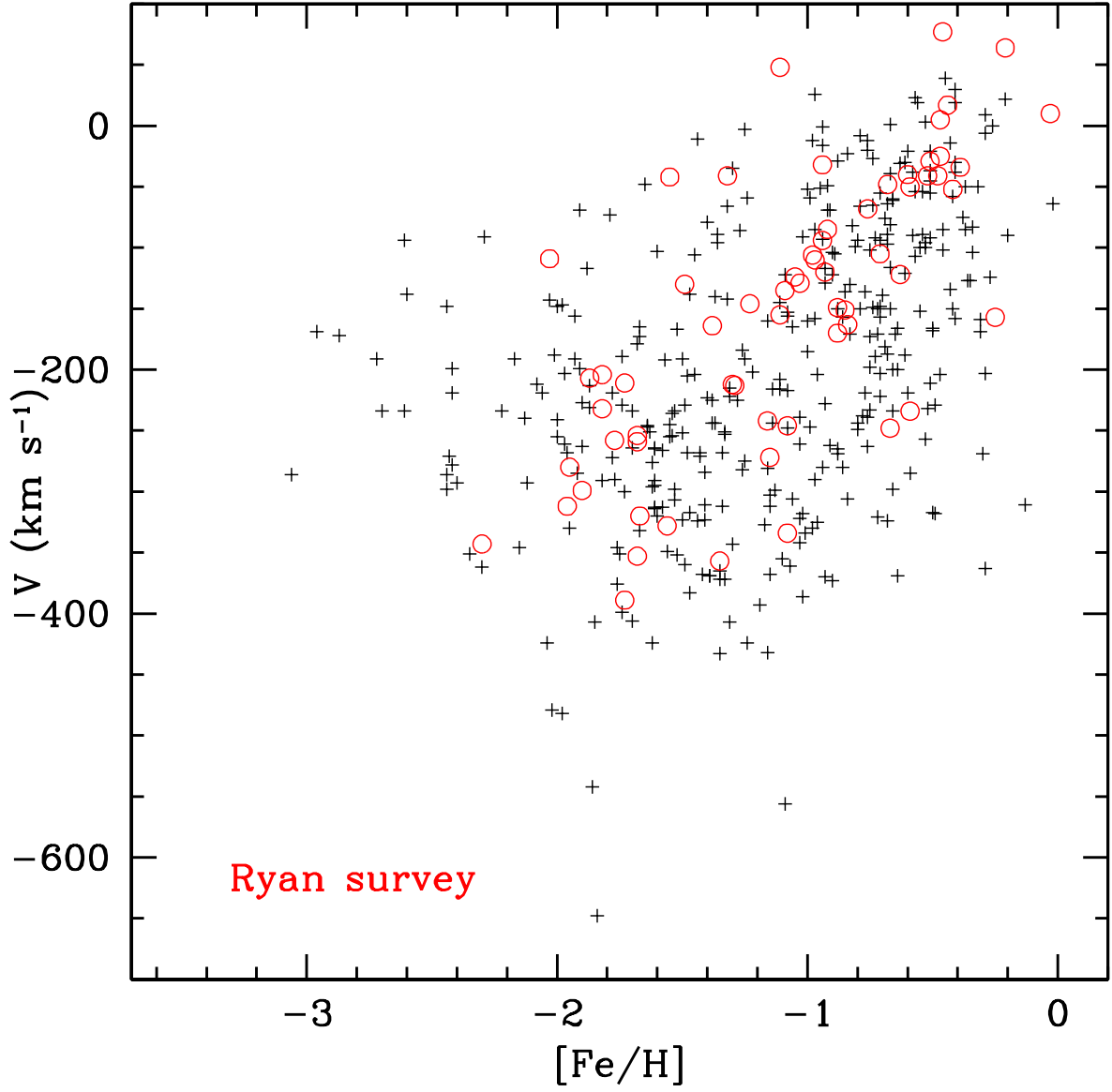}
\caption{(b) The same, but for the Ryan survey.\label{fig1b}}
\end{figure}

\placefigure{fig1b}

\section{RESULTS}

The key finding of this paper is revealed by a careful
inspection of
Figures~1a,b. They suggest that binaries may be less common for
stars on retrograde Galactic orbits. Before discussing possible
explanations for the phenomenon, we begin by exploring the
binary fraction $f$ as a function of metallicity and other
kinematical properties.

\subsection{Binary Fraction as a Function of Metallicity}

In these comparisons, and in all subsequent ones, we have analyzed
the CLLA sample and the Ryan samples separately, and then
the combined sample as well. The comparisons are done
by first ranking each sample according to the variable
being considered, metallicity in this case. Software
divides the samples of single stars into a fixed number of bins 
(chosen to be 10
for the CLLA and combined samples, 5 for the Ryan sample). Bins
differed in size slightly so that they would not overlap
in the values of the parameters. Thus, no bin has any stars
with the same [Fe/H] value as stars in any other bin. The
binary star samples were then sorted according to these
same metallicity ranges, enabling us to calculate the
binary frequency, $f$.
We also computed the average span of the observations (in days)
of the single stars within each bin to check that
we had not introduced
any bias due to greater or lesser time coverage
as a function of metallicity, or any of the various velocity variables
discussed below. (We do not
include binary stars in the time coverage comparisons since observations
of such systems were usually ended when the spectroscopic binary
orbital solution was considered complete.)
No such systematic variations in observing time coverage among the bins
were apparent in any of the comparisons described in this paper.

Figure~2
shows the distribution of $f$ as a function of [Fe/H]. 
Note that
the Ryan sample $f$ values tend to be lower than those of the
CLLA survey. This is a consequence of the smaller number of
radial velocity measurements per star and the smaller span in
the observations. The bins in the CLLA sample have an average
span of 3398 days ($\sigma = 73$ days), while the bins
in the Ryan sample have an average span of only 651 days ($\sigma = 160$ days).
There is a minor difference in $f$ as a function of metallicity,
as seen in Table~1. 
For the combined sample $f = 19 \pm 2$\%
for [Fe/H] $\leq$ $-1.0$ and $f = 24 \pm 2$\% for [Fe/H] $>$ $-1.0$.
If we consider only stars on prograde Galactic orbits (V $\geq -220$ \kms),
the modest difference
in binary fraction 
between the metal-rich and metal-poor stars disappears, as summarized
in Table~1.
The combined sample has $f = 28 \pm 3$\%
for [Fe/H] $\leq$ $-1.0$ and $f = 26 \pm 2$\% for [Fe/H] $>$ $-1.0$.
The conclusion to draw, at least to first order,
is that among high proper motion stars, the binary fraction does
not appear to depend strongly on metallicity. High velocity, metal-poor
halo stars were able to produce binary systems as readily as
low-velocity, metal-rich stars (see also Latham et al.\ 2002).

\begin{figure}
\epsscale{1}
\figurenum{2}
\plotone{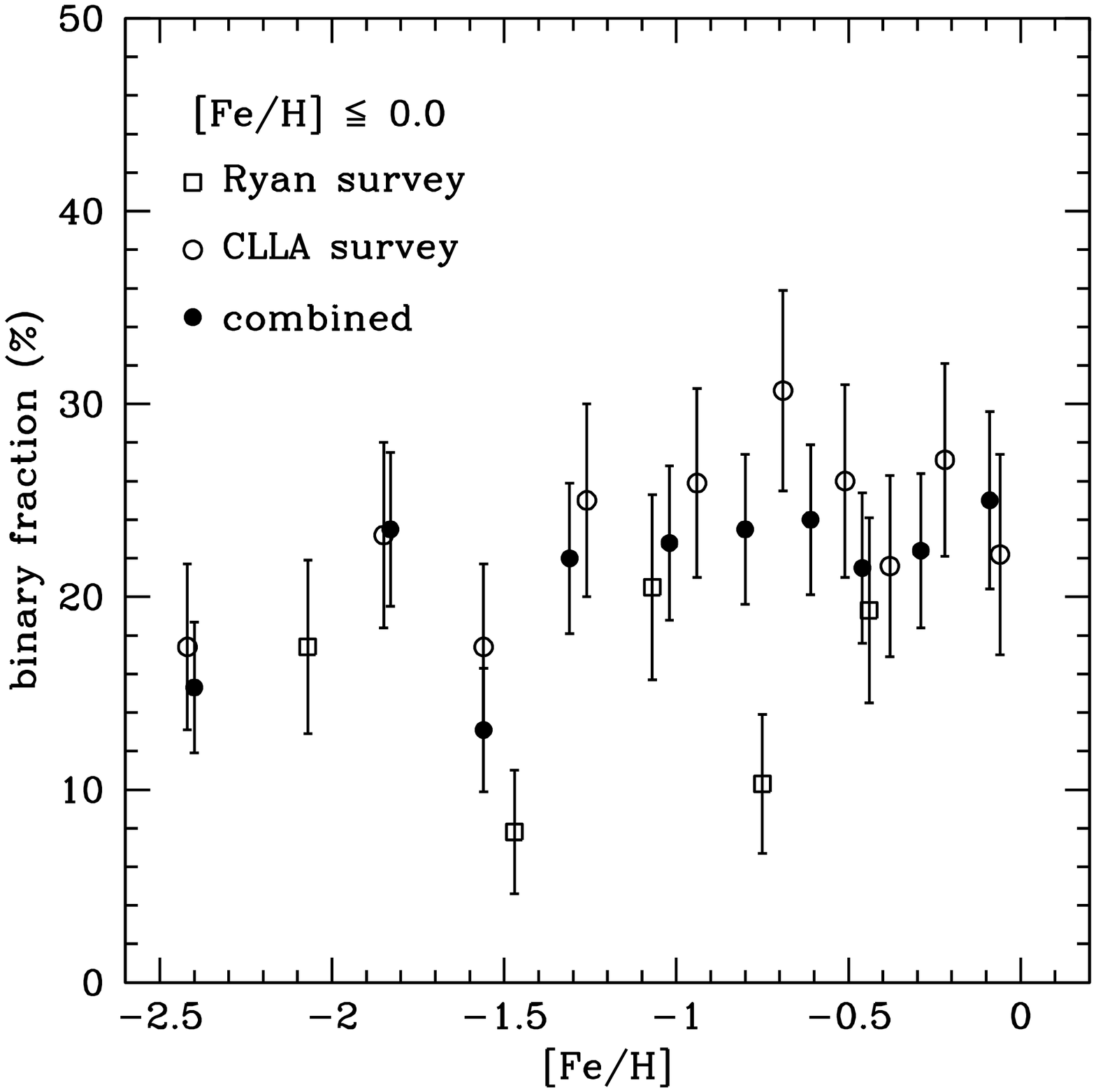}
\caption{The binary fraction
as a function of spectroscopic metallicity. \label{fig2}}
\end{figure}

\placefigure{fig2} 

\placetable{tab1}

\subsection{Binary Fraction of Stars on Retrograde Orbits}

In Figure~3 we revisit the binary fraction as a function of
metallicity, but divide the sample into two parts: prograde (filled circles)
and retrograde (open circles). We have not distinguished between
the halo or Ryan samples here: for clarity we show only the
results for the combined sample.
Two facts are prominent:
(1) The retrograde portion is deficient in binaries relative to
the stars moving on prograde orbits; and (2) Neither the prograde
nor retrograde samples' binary fractions depend on metallicity.

Figures 1a,b show
that roughly half the stars with [Fe/H] $\leq$\ $-1.0$ are moving on
retrograde orbits, while only 6\% and 9\% of the stars in the CLLA
and combined samples with 
$-1.0$ $<$ [Fe/H] $\leq$\ $0.0$ have V $\leq$\ $-220$ \kms.
This low percentage is not, of course, surprising, since it simply
means that for [Fe/H] $> -1.0$, disk stars dominate the sample.
To explore the differences between the prograde and retrograde
stars more carefully, we show in 
Figure~4 the binary fractions as a function of the
V velocity only for stars with [Fe/H] $\leq$\ $-1.0$. The 
Figure shows clearly the deficiency of
binary systems among stars moving on retrograde orbits.
Table~1 quantifies these differences.
The combined sample, with [Fe/H] $\leq$\ $-1.0$ shows a value for $f$
of $28 \pm 3$\% for stars on prograde orbits but only
$13 \pm 2$\% for stars on retrograde orbits. This drops to
only $10 \pm 2$\% for V $\leq\ -300$ \kms. 
Figure~5
shows the cumulative single and binary star
distributions of the combined sample as a function
of V velocity, again for stars with
[Fe/H] $\leq$\ $-1.0$. A Kolmogorov-Smirnov test shows that
we can reject the null hypothesis that both samples
originate from the same parent population
at the 99.998\% confidence level. The 
impressions from Figures 1a,b are confirmed: The local
sample of metal-poor high-velocity stars is deficient
in binaries among stars on retrograde Galactic orbits.

\begin{figure}
\epsscale{1}
\figurenum{3}
\plotone{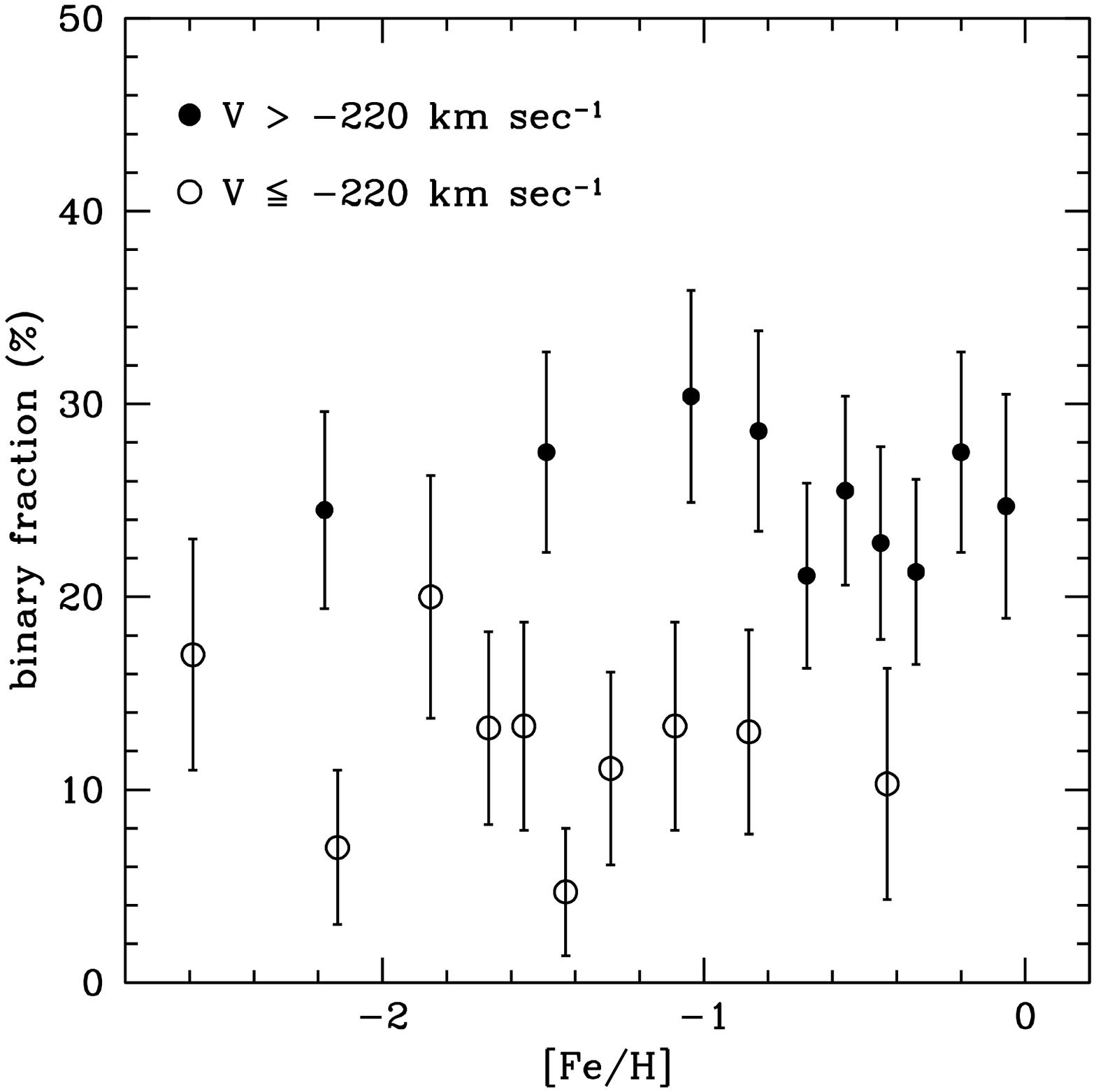}
\caption{The binary fraction
as a function of spectroscopic metallicity
for the combined sample, divided into stars moving
on prograde and retrograde Galactic orbits. \label{fig3}}
\end{figure}

\placefigure{fig3}

\begin{figure}
\epsscale{1}
\figurenum{4}
\plotone{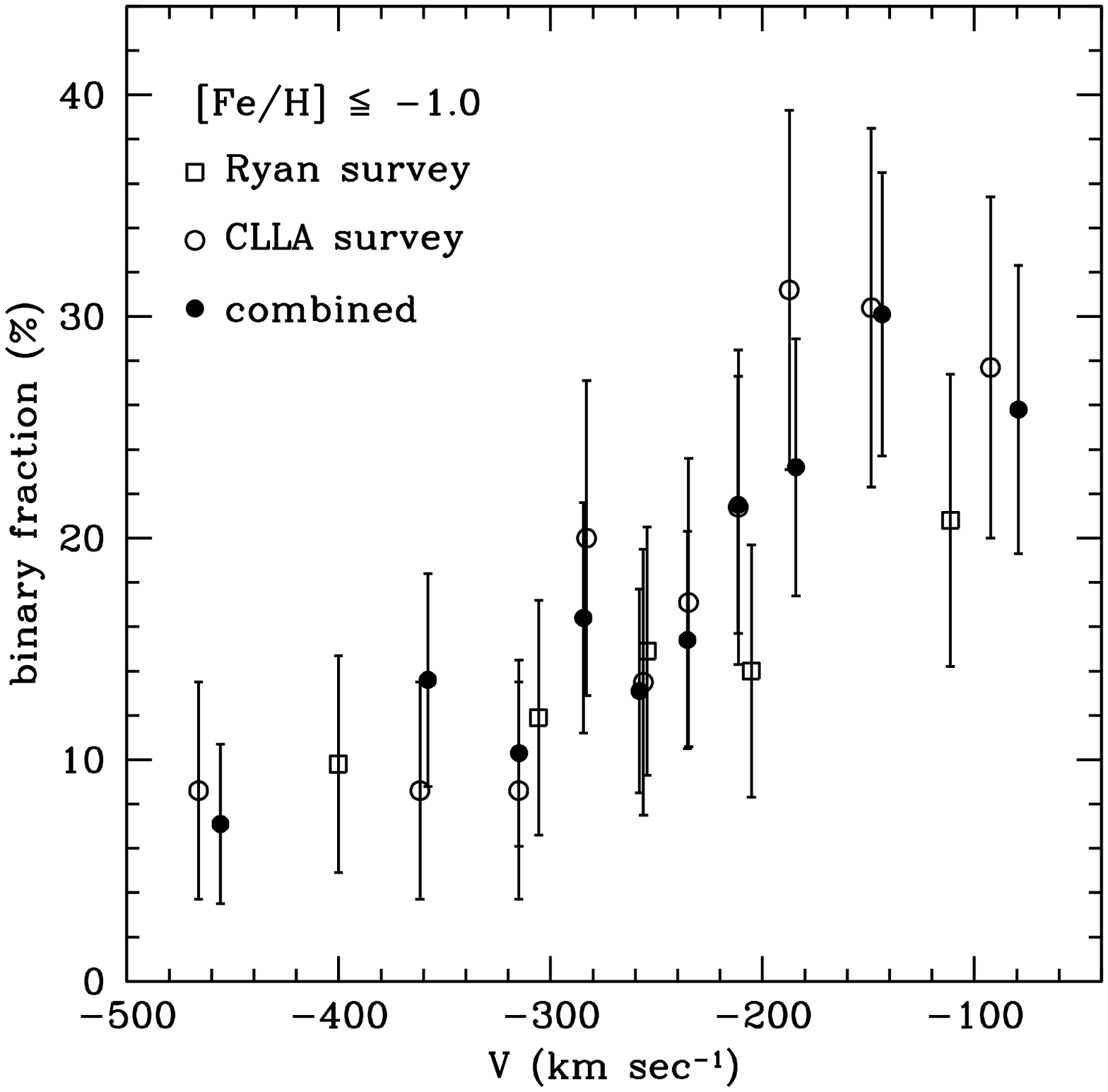}
\caption{The binary fraction
as a function of V velocity for stars with [Fe/H] $\leq$\ $-1.0$.
\label{fig4}}
\end{figure}

\placefigure{fig4}

\begin{figure}
\epsscale{1}
\figurenum{5}
\plotone{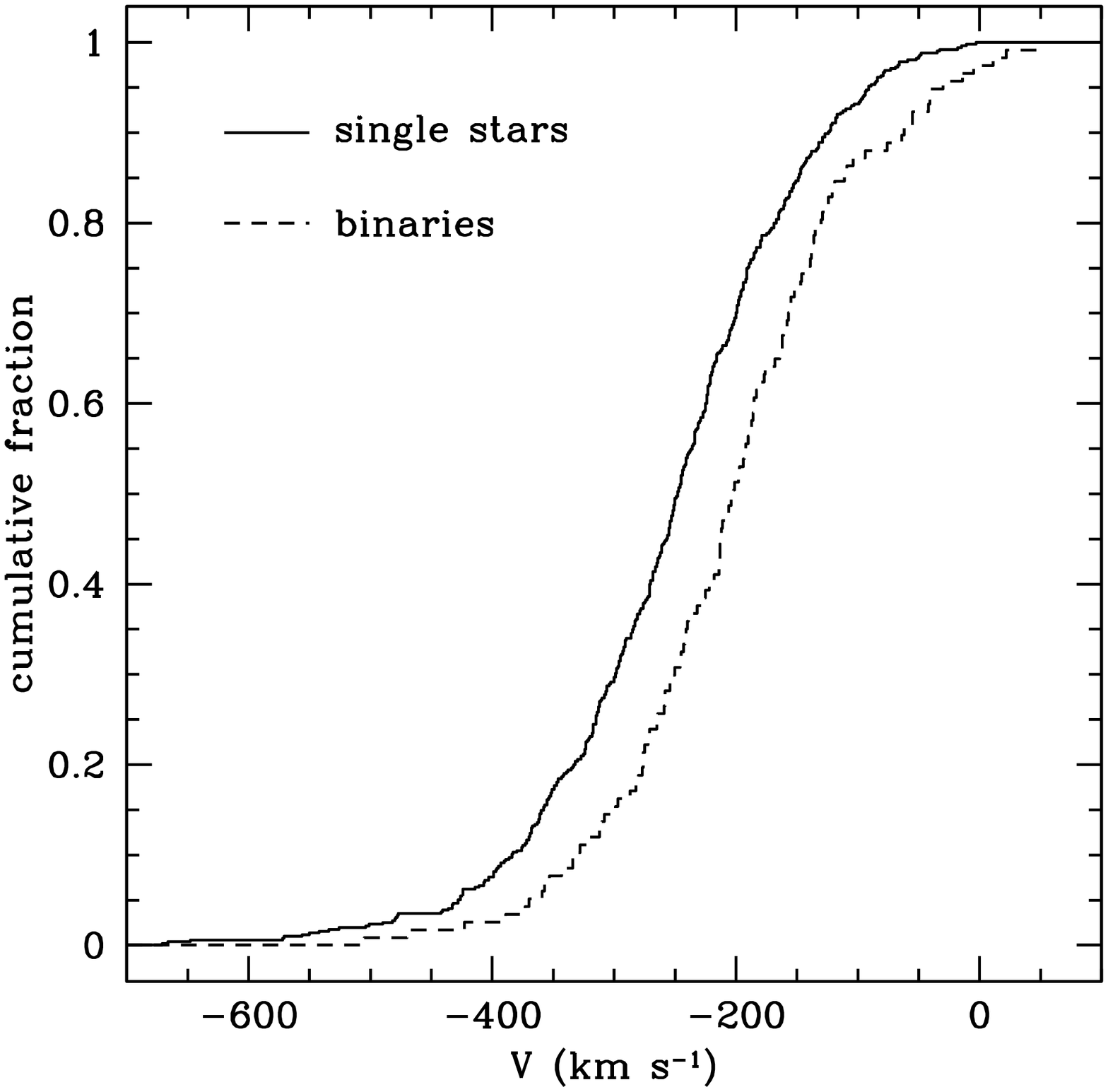}
\caption{The cumulative 
distributions of the single and
binary stars vs. the V velocity. 
The solid and dashed lines
represent the single and binary fractions of the CLLA and combined
samples for the metal-poor stars, with [Fe/H] $\leq$\ $-1.0$.
\label{fig5}}
\end{figure}

\placefigure{fig5}

\subsection{Binary Fraction as Function of Other Velocities}

Figures~6, 7, 8, and 9 show $f$ as a function of U, $|$U$|$,
W, and $|$W$|$. Unlike Figure~4, there do not appear to be
any significant trends in $f$ compared to these velocities,
and simple divisions into low-velocity and high-velocity
bins in Table~1 support this result. 
Kolmogorov-Smirnov
tests of the distributions of these velocities show that the
null hypothesis that, as a function of these velocities,
the idea that the single and binary populations 
are drawn from the same parent populations cannot be rejected
with reasonable levels of confidence.
The only possible effect
is seen in the W velocity. Indeed, for W $\geq$ +100 \kms,
Table 1 shows that $f$ has dropped significantly. 

\begin{figure}
\epsscale{1}
\figurenum{6}
\plotone{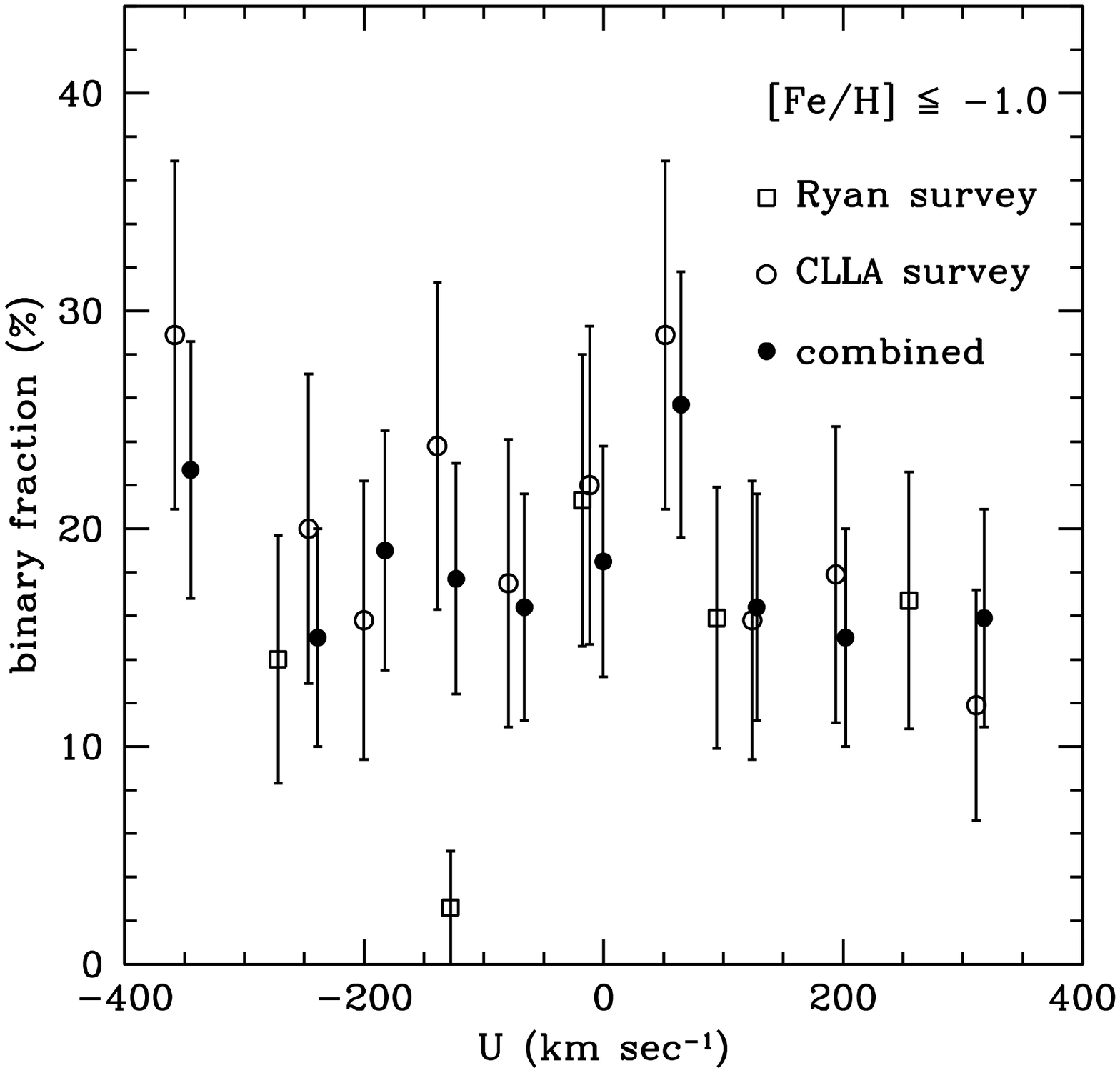}
\caption{The binary fraction
as a function of U velocity for stars with [Fe/H] $\leq$\ $-1.0$.
\label{fig6}}
\end{figure}

\placefigure{fig6}

\begin{figure}
\epsscale{1}
\figurenum{7}
\plotone{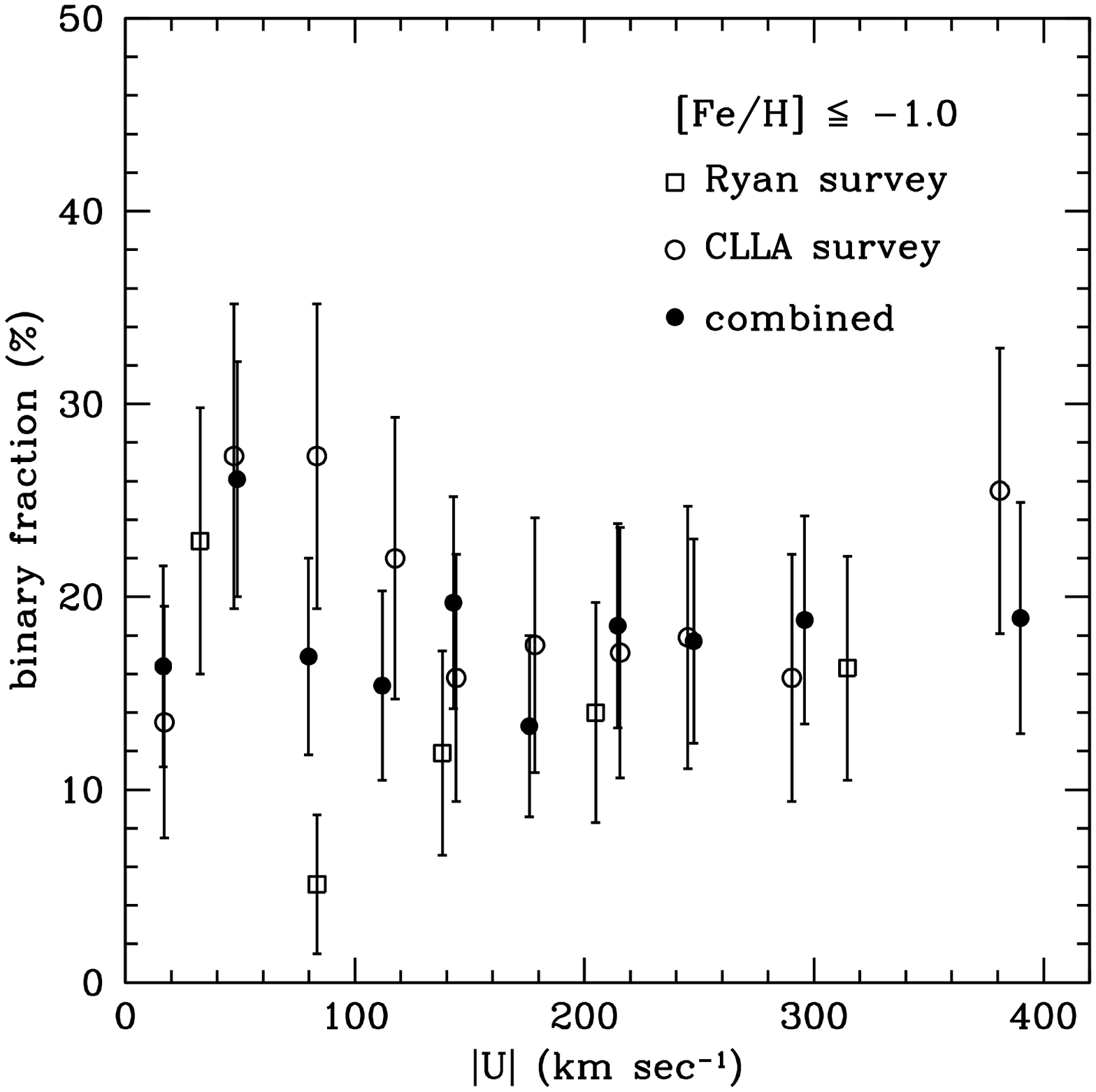}
\caption{The binary fraction
as a function of $|$U$|$ velocity for stars with [Fe/H] $\leq$\ $-1.0$.
\label{fig7}}
\end{figure}

\placefigure{fig7}

\begin{figure}
\epsscale{1}
\figurenum{8}
\plotone{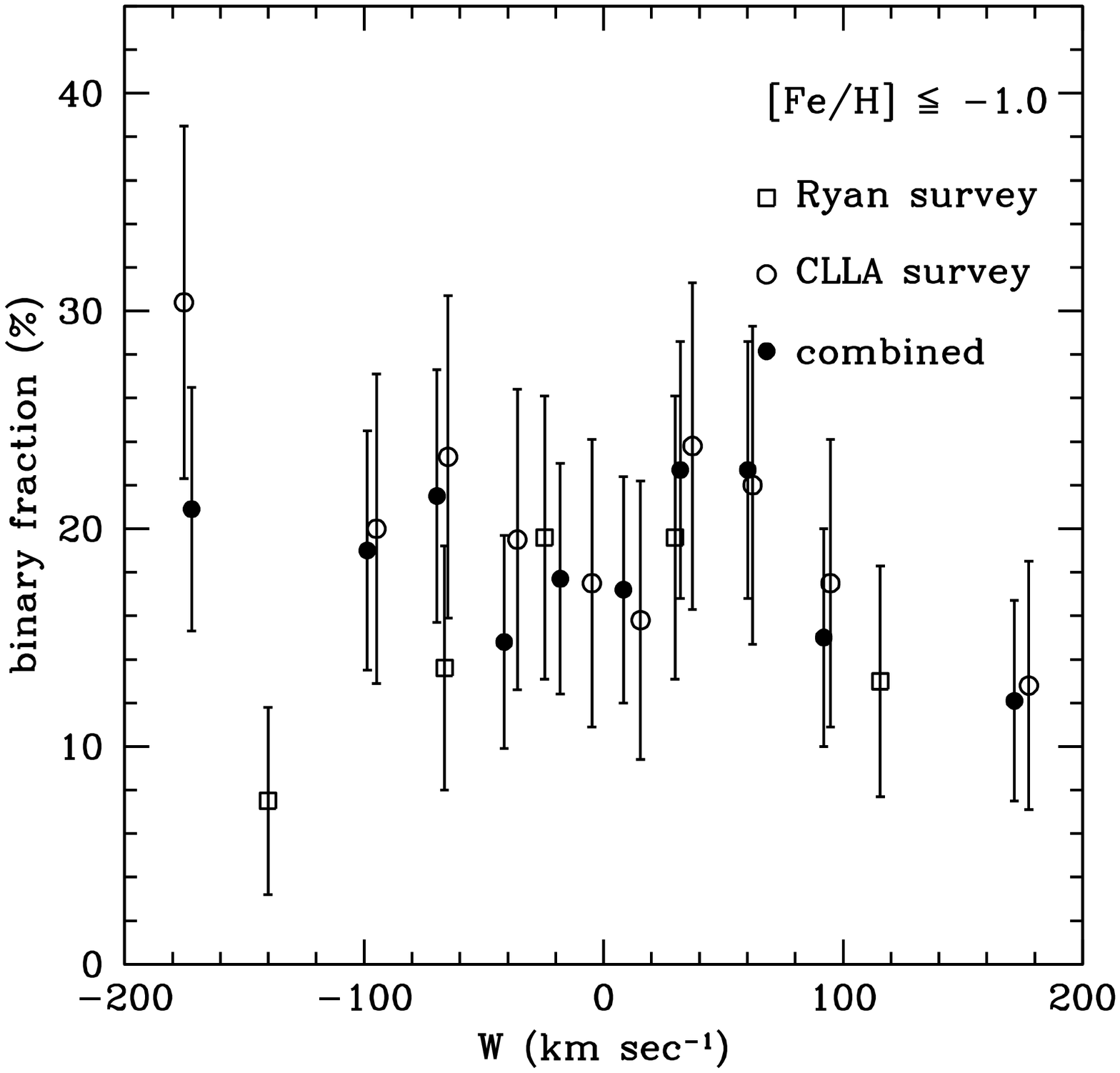}
\caption{The binary fraction
as a function of W velocity for stars with [Fe/H] $\leq$\ $-1.0$.
\label{fig8}}
\end{figure}

\placefigure{fig8}

\begin{figure}
\epsscale{1}
\figurenum{9}
\plotone{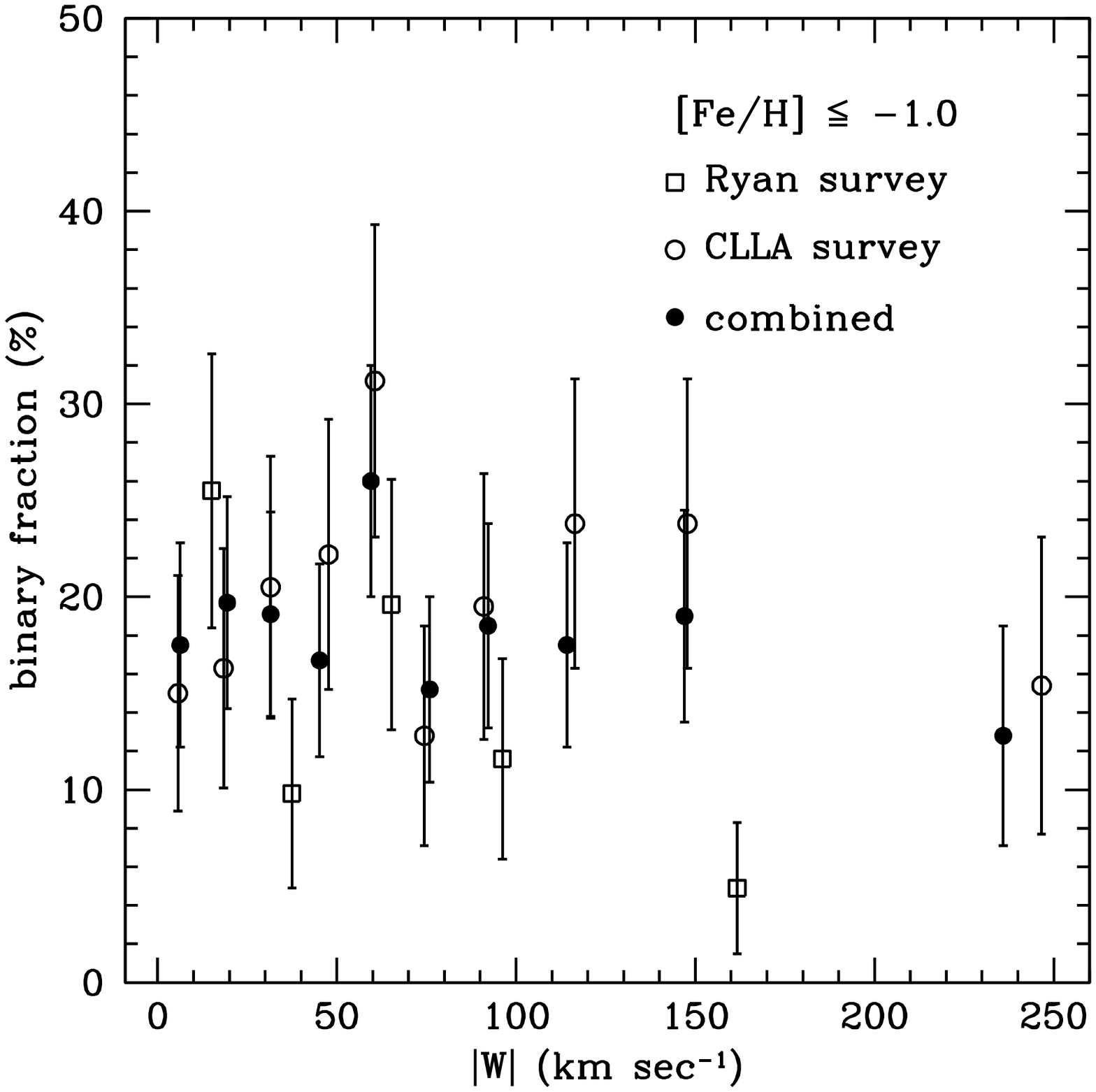}
\caption{The binary fraction
as a function of $|$W$|$ velocity for stars with [Fe/H] $\leq$\ $-1.0$.
\label{fig9}}
\end{figure}

\placefigure{fig9}

\section{THE SEARCH FOR EXPLANATIONS}

It appears, then, that there is a significant deficiency
of binaries among metal-poor stars moving on
retrograde Galactic orbits. What might cause this?

The explanations may be divided into two broad categories.
The ``simplest" class involves the destruction of binaries,
creating the deficit we observe. This affects a significant
fraction of the stars moving on retrograde orbits. Consider,
for example, the 170 stars with V $\leq\ -300$ \kms. The typical
binary fraction in our sample of about 25\% predicts that
about 40 or more binaries should be found, while we detected
only 17. For V $\leq\ -220$ \kms, there are 374 stars, so
over 90 binaries are expected, and only 47 are found. {\em It appears
that of order half of the binary systems on retrograde orbits would have
to have been disrupted. That is about 10-15\% of the total number of
stars, so the effect is indeed significant.}

The second broad class of explanations is that instead of
binaries being disrupted, a mechanism exists to add significant
numbers of single stars, thereby diluting the binary fraction. 
Consider again the same sample with V $\leq -300$ \kms. Normally,
17 binaries would be found among a total population of only about 70
stars, not the 170 stars in our subsample. The dilution of the
sample by ``extra" single stars is thus
very large: roughly 100 stars, or over 60\% of the
observed sample! For V $\leq\ -220$ \kms, the 47 binaries should be found
among a sample of 188 stars, not the much larger sample of
47 binaries and 327
single stars. In this case, there are 186 too many single stars.
{\em Both of these retrograde samples 
imply that of order 50\% of the total samples would have to have been
``added" as single stars.} 

Given the relative magnitude of the number of single stars to be
added ($\approx 50$\% of the total sample) 
compared to the number of binary systems to 
be disrupted ($\approx 50$\%, but only about 10\% of the total sample),
it appears that the latter 
is a more plausible explanation than
dilution by single stars. We therefore consider first
whether binary disruption is a viable explanation.

\section{DISRUPTION OF FIELD BINARY STARS WITHIN THE GALAXY}

Binary systems may be disrupted by close encounters with compact or extended
massive objects, and somehow the destruction 
mechanism must be enhanced for binary
stars moving on retrograde Galactic orbits. 
We approach this topic from both theoretical and
observational perspectives.

\subsection{Theoretical considerations for tidal disruption}

We ask if massive objects within the Galactic plane could be especially
effective in disrupting those halo binaries that are primarily on retrograde
orbits in the Galaxy. One differentiating characteristic for these stars is
the larger relative velocity with which they encounter other objects in the
disk. Although a larger relative velocity and shorter
interaction timescale, such as between
objects on prograde and retrograde orbits, generally implies a less
destructive encounter, a
larger relative velocity can drive encounters into the impulsive regime
where collisions are most destructive. Binaries on prograde orbits would
experience encounters at lower relative speeds, where the encounter may
be adiabatic and the binary would be less vulnerable to disruption. The
outcome of such encounters thus may depend on whether they are
in the adiabatic or impulsive regimes.

A simple criterion for the impulsive regime is that the ratio of
the encounter time to the binary orbital period $P$, should be less than one
(Spitzer 1958):
\begin{equation}
\label{eq:impulse}
{{(2b/v_{\rm rel})}\over P} < 1,
\end{equation}
where we have taken the encounter time to be 
the interval that the stars take to cover a
length equal to twice the collision impact parameter $b$, at the relative
velocity, $v_{rel}$. This condition means that, for the encounter to be
impulsive, the impact parameter should be smaller than
\begin{equation}
\label{eq:timing}
\left({{b}\over{1\, AU}}\right) < 10.5\, \left({{P}\over{\rm 1~yr}}\right)\,
\left({{v_{\rm rel}}\over{10^2\, {\rm km~s^{-1}}}}\right).
\end{equation}
An extreme case occurs if
we take $v_{\rm rel} = 400$ \kms. The encounter would
have to be within $0.2$ pc to be impulsive for a
binary with an orbital period as long as
$10^3$ years. This
is a much longer period
than those of the spectroscopic binaries
that dominate the binary statistics in our study.
Any shorter period binary would be in the adiabatic regime, where
the energy transfer from the encounter to the binary becomes very inefficient.
Since we see the binary deficiency among a sample of stars whose orbital
periods are less than ten years, the relevant encounters must have
been much closer, 0.002~pc, to enable impulsive encounters that
would disrupt the binary systems.

Even encounters with massive objects, like giant molecular clouds, would
affect only the widest binaries. Although the velocity kick
produced by an impulsive encounter increases linearly with the mass of
the perturbing object, from the impulsive condition, it is clear
that an encounter with an extended object as large as a molecular cloud
can not be close enough to be in the impulsive regime. The energy transfer
in the impulsive regime quickly approaches zero
as the physical size of the perturber increases, thus canceling
the effect of the large perturber mass. Weinberg, Shapiro and
Wasserman (1987) have investigated this phenomenon in detail
and concluded that binaries with separations of 0.1~pc or less
can survive for $10^9$ years when subjected to perturbations by
stars and molecular clouds, if confined within the thin
galactic disk, a condition unlikely to be met by the high-velocity retrograde
stars in our sample. Thus tidal disruption due to encounters with stars
and molecular clouds is not able to pack enough energy within the tight
confines of the binaries in our sample to be able to explain the effect
seen in Figure~4. 

What about black holes?
The binaries most readily
disrupted are those with the lowest binding energies, meaning the widest
separations. The widest known pairs in the solar neighborhood have
projected separations of about 0.1 pc (Latham et al.\ 1984). Indeed,
the widest pair in our samples, HD~134439 and HD~134440,
has a projected separation of just under
0.05 pc. As Appendix~A discusses, and Figure~A-1 shows, binary
separations this large imply large values of the impact parameter for
massive point particle perturbers, 
and, therefore, only a modest density of perturbers is required. However,
to disrupt the spectroscopic binaries in our sample, with periods of
order a decade or less, and separations of a few AU, the impact
parameter for encounters becomes extremely small, 0.001~pc or less,
and the implied number density of perturbing black holes would be
totally out of proportion compared to the
inferred mass density in the Galactic plane (see Bahcall, Flynn, \&
Gould 1992). 

\subsection{Observational tests}

We may also rule out lower velocity encounters with massive objects
confined to the disk from a consideration of Figures~8 and 9. The
stars in our sample with the $|$W$|$ velocities closer to zero spend more
of their time near the plane, and no binary deficiency is
seen at such velocities.

The Galaxy's central mass concentration has been claimed to be an
effective environment for shredding star clusters (Aguilar, Hut,
\& Ostriker 1988). Could it also be efficient at destroying stellar
binary systems? While the stars in our sample all lie
within a few hundred parsecs of the Sun, in some cases
the calculated perigalacticon distances are very small.
This is strongly correlated with the Galactical orbital
angular momentum, such that stars with V $\approx -\Theta_{0}$
(and hence low angular momentum) fall towards the Galactic
center. 
The commonly-adopted
value for $\Theta_{0}$, the circular velocity
at the Sun's Galactocentric distance, is 220 \kms\ (Kerr \& Lynden-Bell 1986),
although there is some evidence for an even higher value
of 270 \kms\ (M\'{e}ndez et al.\ 2000). If the Galactic
center plays a major role in binary disruption among
our survey's stars, we would expect to see a minimum
in the binary frequency near V $\approx$ $-220$ (or $-270$)
\kms, with a much-reduced disruption effect and therefore
higher binary frequency at {\em both} lower and higher
V velocities (i.e., for stars with higher levels but
opposite signs of angular momentum). Figure~4 is
not consistent with this model. On the other hand, the V velocity is only
a component of the total velocity in the solar neighborhood
directed perpendicular
to the direction toward the Galactic center. We therefore
consider the combination of the V and W velocities. Figure~10
shows the binary percentages 
for the ``VW" velocity, 
\begin{equation}
VW = [(V+\Theta_{0})^{2} + W^{2}]^{1/2}.
\end{equation}
Stars with very small values of VW pass close to the
Galactic center, but the binary percentage remains
constant, again indicating that the Galactic center
is not responsible for tidally shredding most of the original
binaries that have become single stars in our sample.

\begin{figure}
\epsscale{1}
\figurenum{10}
\plotone{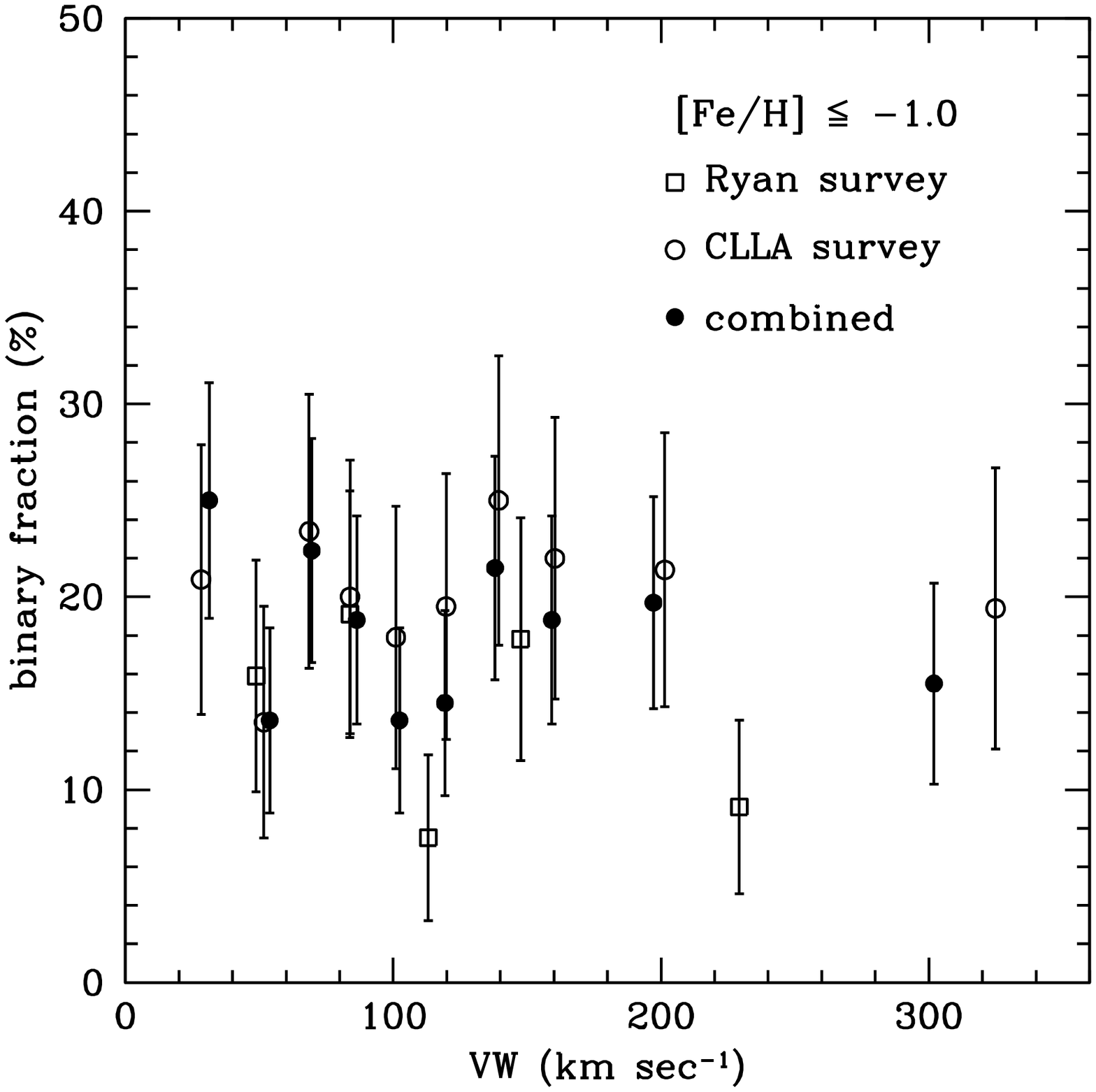}
\caption{The binary fraction
as a function of ``VW" velocity for stars with [Fe/H] $\leq$\ $-1.0$.
\label{fig10}}
\end{figure}

\placefigure{fig10}

Finally, there is additional evidence that tidal disruption is
not the primary explanation. We have shown that tidal effects
are stronger for wider, less tightly bound,
and longer-period systems, as discussed above.
If tidal effects are the
explanation of the binary deficiency seen in Figure~4,
we expect that the surviving binaries would, on average,
have shorter periods than for samples that have apparently
not been subject to such disruptions. We therefore
must look closely at the binaries at both ends of
the V velocity distributions and compare the period
distributions. We rely here only upon the CLLA sample
since its time coverage is much longer and the
numbers of observations have enabled us to obtain more
spectroscopic binary orbital solutions 
than in the case of the Ryan sample. Specifically,
we consider
the three lowest and highest V velocity bins of our
CLLA sample as shown in Figure~4, stars with V $\leq$\ $-300$ \kms,
and V $\geq$ $-200$ \kms, respectively, and to which we
now refer as the retrograde and prograde binary samples.
We consider first the systems
whose periods are too long for us to determine:
common proper motion pairs and visual binaries.
Among the 9 binaries in the retrograde sample, three
($33 \pm 19$\%) fall into this long period regime.
In comparson, 11 of
the 46 binary stars in the prograde sample ($24 \pm 7$\%) fall into the
long period domain. We note that one star in the retrograde sample
and two stars in the prograde sample are both spectroscopic
binaries as well as members of common proper motion pairs, but for the
purposes of this part of our discussion, all we really wish to
see if there is an absence of long-period systems among stars
on retrograde vs.\ prograde orbits. We do not.
We now consider the spectroscopic binary orbital periods (including
preliminary periods) in the retrograde and prograde samples to see
if the former show a dearth of long-period systems.
Figure~11 shows the
period distributions for the 7 retrograde and 35 prograde
binaries for which we have determined or estimated periods.
The distributions are very similar, and the Kolmogorov-Smirnov test
shows that we cannot reject the hypothesis that they are drawn from the
same populations.

\begin{figure}
\epsscale{1}
\figurenum{11}
\plotone{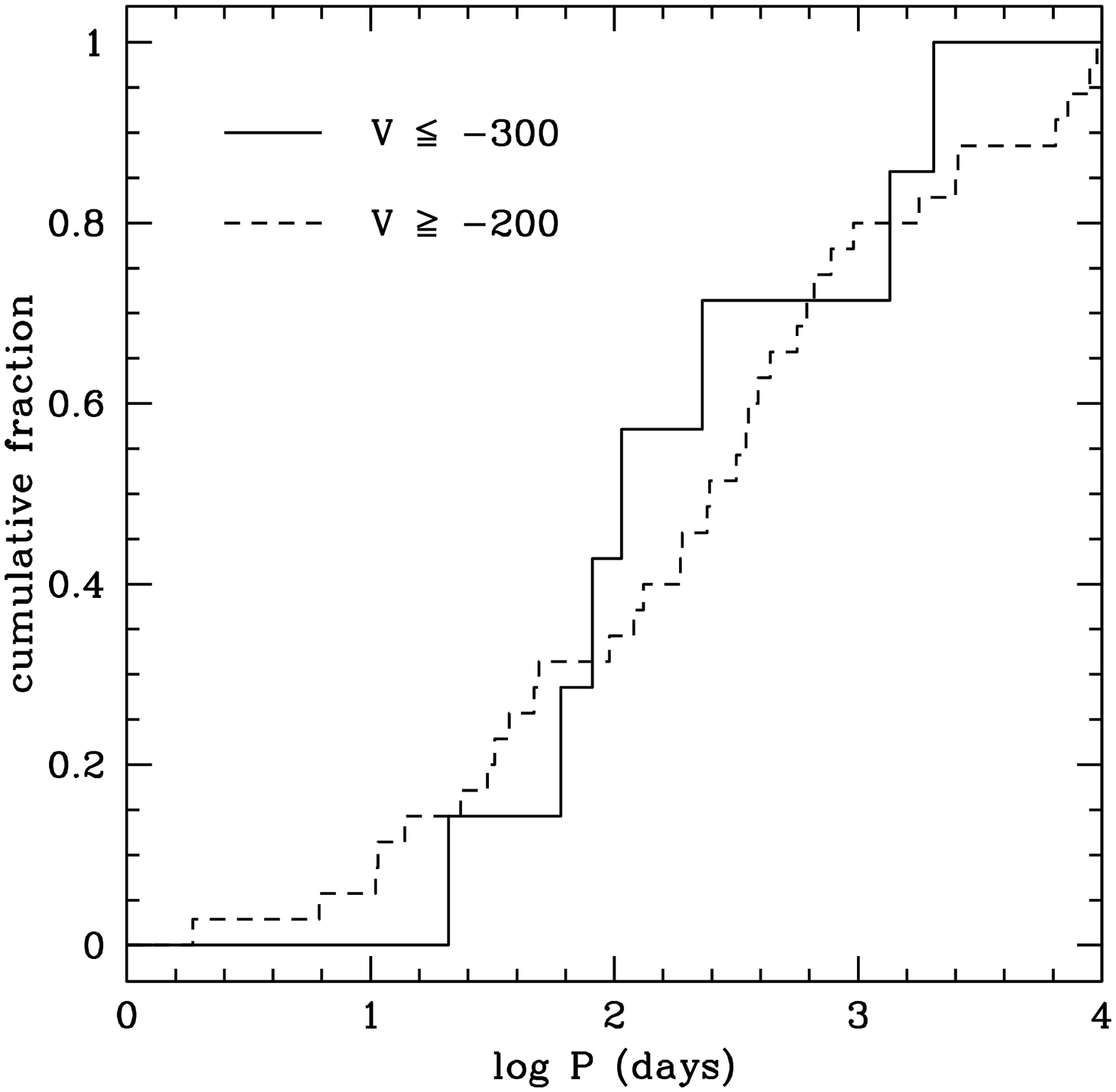}
\caption{The cumulative period distributions
of the 7 binaries with known periods and V $\leq\ -300$ \kms\
compared to those of binaries with V $\geq\ -200$ \kms.
\label{fig11}}
\end{figure}

\placefigure{fig11}

We conclude that in our binary-poor retrograde sample 
there does not appear to be any preference
for short-period ``survivors" vs. long-period binaries
predicted by tidal-driven disruption mechanisms.

No binary disruption mechanism appears to be capable
of explaining the paucity of binaries moving on retrograde
Galactic orbits, and we are forced to explore possible sources of
excess numbers of single stars even though, as we have seen, we
will need to identify a source or sources of single stars for
almost 50\% of the stars moving on retrograde Galactic orbits.

\section{GLOBULAR CLUSTERS AS SOURCES OF SINGLE STARS}

We explore here the two options that rely on globular clusters and which
could provide single stars while sequestering binary systems.
In the first case we assume that a binary-single star encounter
within the cluster leads to a more tightly bound binary
system and the ejection of a single star. In the second,
we assume that dissolved (or dissolving) clusters have primarily populated
their outer regions with single stars, thereby increasing
their numbers in the field, and decreasing the binary fraction, at
least in restricted regions of velocity space.

\subsection{Ejection of single stars}

To explain Figure~4 by binary disruption within Galactic
globular clusters
would require that the stars on the most extreme
retrograde Galactic orbits acquired a ``velocity kick"
during a binary disruption. The average metal-poor
halo star or cluster has a mean V velocity of $\approx -200$ km/s,
and the one-dimensional velocity dispersion of the field
stars and the ensemble of metal-poor globular clusters is 
of order 100 \kms. To introduce
the sort of asymmetry seen in Figure~4 therefore requires
that the kick should add about the magnitude of the halo's velocity dispersion
to the stars' space motions. Qualitatively, the idea has some
merit, at least for the V velocity. Single stars ejected
toward more negative V velocities would provide an excess
of single stars, which we will have interpreted as a deficiency
of binaries. Because stars ejected toward more positive V velocities
will have space velocities
closer to the Local Standard of Rest, they would be less likely
to appear in samples, like ours, that are selected from proper
motion catalogs.

Quantitatively, this explanation has problems. First,
the original binary periods
would have been on the order of days ($M_{1} + M_{2} = 2~M_{\odot}$
and $v_{orb} = 100$ km/s leads to 
P$_{orb} \approx 20$ days, with P$_{orb} \propto M/v^{3}$).
It is unlikely that there were enough such short-period halo binaries
available for disruption. More important, however,
the ejection should occur over 4$\pi$ steradians, and so it
should be manifested in the U and W velocity directions as well.
Figures 6-9 rule out this explanation.

\subsection{Globular cluster dissolution?}

Here we ask if the low fraction of binary systems on
retrograde orbits might have been caused by the evaporation
of stars from a globular cluster or clusters moving on retrograde orbit(s).
Dynamical evolution of a globular cluster, for example,
leads to a greater central concentration of tight binary
systems and a more extended distribution of lighter single
stars (see Hut et al.\ 1992 for an excellent discussion). 
Many of the original globular clusters in our Galaxy
may have undergone such evolution, to the point of
much reduced sizes or total evaporation (Aguilar 1993). There are several
lines of evidence that support the idea that many
of the Galaxy's original clusters have suffered such
a fate. First, Chernoff \& Djorgovski (1989) found that
clusters within the central 3 kpc of the Galaxy are more concentrated
and more likely to show central power-law density profiles, indicative
of post-core collapse evolution, and in agreement with the cluster
evolution models of Chernoff, Kochanek, \& Shapiro (1986). Second, the
selective erosion of clusters in radially plunging Galactic orbits
can explain the observed difference in kinematics between the stellar
halo and globular clusters (Aguilar, Hut, \& Ostriker 1988).
Third, studies of young stellar systems with
masses like those of globular clusters in starburst
galaxies and merging galaxies 
have revealed that the mass spectrum
follows a power law
(see Whitmore 2003
for a review; also Whitmore et al.\ 1999;
Zhang \& Fall 1999).
This is completely unlike that
of the current Galactic globular clusters, which has
a Gaussian-like mass distribution (Harris 1991). One way
to evolve from a power law distribution to that observed
today is through more rapid dissolution of lower mass
systems (see Fall \& Zhang 2001; Vesperini \& Zepf 2003).
Finally, we can actually see cluster dissolution happening through detection of 
cluster tidal tails (Meylan \& Heggie 1997; Leon, Meylan, \& Combes
2000; Odenkirchen et al.\ 2001, 2003; Siegel et al.\ 2001).
We recognize the difficulty, however, of relying on the dissolution
of many clusters to provide the excess of single stars, because
this mechanism then also requires a means to favor dissolution
of clusters moving on retrograde rather than prograde orbits.

We approach this issue from two perspectives, asking first
if there is a difference in terms of dynamical evolutionary
state for clusters moving on prograde or retrograde orbits,
and then seeking signs for dissolved clusters in the form of
``moving streams" in velocity space. Detection of such streams would have
relevance to the search for dark matter substructure
within the Galactic halo (Ibata et al.\ 2002) since significant
departures from a smooth mass distribution (i.e., substructure)
enhances dissolution of streams, their
detection may provide a limit to the degree of
the substructure in the Galactic halo. 

We have used the same database as Chernoff \& Djorgovski (1989)
to evaluate the dynamical evolutionary state of globular
clusters. The two relevant parameters are the concentration
class, $c$, and the type of model applicable to the stellar
distributions. Specifically, which clusters show evidence
for post-core collapse? Because Chernoff \& Djorgovski (1989)
found clear evidence favoring enhanced dynamical evolution
for clusters closer to the Galactic center, we limit
our study to those clusters which are not only outside
some limiting Galactocentric distance, but also restrict
our study to only those clusters whose Galactic orbits 
do not carry them closer than a certain distance, using
the orbits calculated by Dinescu, Girard, and van~Altena (1999). 
For the 19 clusters (excluding the very distant cluster
Palomar~3)
with perigalacticon values exceeding 3 kpc, five are
classified as ``PCC" or ``PCC?", and four of those are on
prograde orbits. The mean concentration class of the nine
clusters on prograde orbits is $<c> = 1.84 \pm 0.09$ ($\sigma = 0.28$),
while that of the four clusters on retrograde orbits is
$<c> = 1.41 \pm 0.12$ ($\sigma = 0.24$). If we move the
perigalacticon limit out to a distance of 5 kpc,
there is only one ``PCC" cluster, NGC~7078 (M15), and it is
moving on a prograde orbit. The same four clusters define the
retrograde sample, but the prograde sample has declined from
nine to five clusters, and $<c> = 1.85 \pm 0.11$ ($\sigma = 0.25$).
There is certainly no evidence for enhanced dynamical evolution
of globular clusters moving on retrograde orbits. Indeed, if
anything, the evidence favors enhanced evolution for
clusters moving on prograde orbits.

We turn now to the search for star streams.
No globular clusters are close enough themselves
for their stars to be seen in our survey, but moving
streams occupy larger volumes than do individual clusters,
and while the space densities of stars in streams are
low, their densities within velocity space may lead to
their detection, even in our modest sample. Because
the physical volume of our survey is small, detection
of stars in a stream requires that each star have very similar
space velocities. Thus the U, V, and W velocities should
be quite similar, consistent basically only with observational
uncertainties. With the exception of $\omega$~Cen,
stars in any one globular cluster have very small
(e.g., Langer et al.\ 1998) to non-measurable differences
in [Fe/H] (Suntzeff 1993). We therefore expect 
any candidate stream to
involve stars with very similar metallicities
if it originated from a dissolved or dissolving
globular cluster. We therefore 
conduct a search for dissolved clusters by confining searches
to narrow windows of metallicity as well as of velocities.

How many streams might we expect?
Our full sample of
stars with reliable metallicities includes 1638 stars,
with an average distance of 162~pc, and 90\% of them
lying within 330~pc. The metal-poor, high-velocity
stars may be seen to greater distances due to their
higher tangential velocities. The 423 stars
with [Fe/H] $\leq\ -1.0$ and V $\leq\ -220$ \kms\ have an
average distance of 286~pc, and 90\% of them lie within 470~pc,
a significantly larger volume. For V $\leq\ -300$ \kms,
the average distance increases to 325~pc, and the 90\%
distance increases to 520 pc.
Our maximal sample volume is
thus about $6 \times\ 10^{8}$ pc$^{3}$ (and is much smaller
for slower-moving objects). 
The two nearest globular
clusters are M4 and NGC~6397, with distances of about 2~kpc, and
have proper motions of about 0.02\arcsec\ per year
(Dinescu et al.\ 1999), corresponding
to tangential velocities of about 200 \kms. For these clusters'
stars to have proper motions large enough to have been detectable
in our sample, with a proper motion limit of $\mu > 0.2$\arcsec\ per year,
the clusters' stars would have to be ten times closer, or about 200~pc.
There are too few clusters in the Galaxy now for us to
have one so nearby. But if most of the Galactic halo's field
stars came from globular clusters, then since
the halo field stars' combined mass exceeds that in the
current population of globular clusters by a factor
of about one hundred, there may have been of order a hundred times
as many clusters as now exist 
(or more, given dissolved
clusters' lower masses). Therefore 
the typical distance
from the Sun may have been (100)$^{\frac{1}{3}}$ smaller
than it is now, and the nearest cluster would have been
only 400~pc away, and thus we might expect to see evidence for its
remnants in the form of a star stream.

{\em However, dissolved clusters occupy much larger
volumes than bound clusters, and so would be even nearer to the
Sun, on average.} And, as we noted above, while the spatial densities of stars
might be low, they might be discernable in velocity space.
A factor of roughly one hundred increase
in spatial number density of dissolved clusters
means that the proximity of a stream to the
Sun is increased by another factor of (100)$^{\frac{1}{3}}$, and
hence there may well be several to many
streams from dissolved or dissolving
clusters in the solar neighborhood. How many there are
depends on how uniformly the stars are dispersed along the
clusters' orbits and, of course, how many clusters originally
were formed in the Galactic halo. If most of these
``dissolved" clusters actually take the form of lighter
single stars dispersed along the orbit plus an undetectable
remnant of tightly packed binary stars and binary stellar
remnants, we might be able to explain our observation of
a deficiency of binary stars on retrograde orbits. We would, of
course, still have to explain why dissolved clusters should
be found preferentially on retrograde orbits. Or we could
invoke a single cluster, whose binary-rich remnant is
beyond our view, and whose single stars populate a significant
fraction of the metal-poor stars in the solar neighborhood.

How many stars from an
individual cluster might we expect, assuming its orbital
stream passes through the local volume?
Let's assume that the cluster stream's stars
have diffused uniformly along the
orbit, but perpendicular to the stream by an amount considerably
smaller than the diameter of our local sampling volume. In that
case the number of detectable stars would be related to the number in 
the original cluster, $N_{\rm stars}$, and the ratio
of the length of the orbit within the local volume divided
by the total orbital length. Assuming a circular orbit,
the number of detectable stars would be
\begin{equation}
\label{eq:stream}
N_{\rm det} \approx N_{\rm stars} d_{\rm local}/ (2\pi R_{0}),
\end{equation}
where $d_{\rm local}$ is the thickness
of the sample volume and $R_{0}$ is the distance of the Sun from the
Galactic center, 8.0~kpc. Thus for $N_{\rm stars}$ = $10^{5}$
and $d_{\rm local}$ = 300~pc, we might find as many as
a thousand stars, with longer orbits leading to fewer
stars in our search volume. And notice that the odds are somewhat
improved for stars on retrograde orbits since the effective
local sample volume increases in size.

Our actual net, however, would be far less than a thousand
stars, for two reasons. First, the least massive globular
clusters would have dissolved most rapidly, and the original cluster
might have included only $10^{3}$ stars rather than $10^{5}$.
Second, our surveys cover only a limited range
of temperature or color, and hence a limited range
of absolute magnitude. Hence the surveys cover 
only a small fraction of any
cluster's stars. Consider the luminosity function of the globular
cluster M13 (Richer \& Fahlman 1986) shown in Table~2.
If we consider only those stars
in our surveys with metallicites between [Fe/H] = $-1.3$ and
$-1.7$ (compared to that of M13, with [Fe/H] = $-1.5$), we
find a total of 2, 94, 87, and 42 stars in the color
ranges that correspond to the absolute and apparent
magnitudes of the M13 luminosity function. Notice that our
survey luminosity function is dropping at redder colors
(and fainter absolute magnitudes) than is the M13 luminosity
function. In the final column of Table~2 we assume that
the color range corresponding to ($B-V$)$_{0}$ = 0.414 to
0.531 is complete in our surveys. We normalize the other
magnitude and color bins by the ratio
of 43.3/94 times the number of stars in our surveys.
The total number of stars in our survey is then 103.6,
compared to the 973.6 in M13. In other words, the color
and magnitude selection criteria in our surveys mean
that our sample is likely to include, at most, only 10\%
of any dissolved cluster's main sequence stars. This
is of course an upper limit since it is unlikely that
our surveys would find all of a dissolved cluster's
stars in our search volume in the ($B-V$)$_{0}$ = 0.414 to 0.531
magnitude range. And the M13 luminosity function is itself
incomplete at the faintest end.
We conclude that a completely dissolved globular cluster whose
orbital stream passes near the Sun and whose stars are moving
at high velocities relative to the Sun might yield only a
few stars.

\placetable{tab2}

If dissolved or dissolving clusters is the explanation
for the retrograde binary deficiency, we have several new questions to answer.
First, do we see any signs of streams?
While streams may be disrupted, their
detection would lend support to this hypothesis.
Second, if they are detected, are they moving primarily on retrograde orbits?
If multiple dissolved clusters are the cause of the effect, we then
also require an enhancement effect to preferentially populate clusters
moving on retrograde Galactic orbits.
On the other hand, if only a single
large object which had a retrograde orbit is involved,
are there other clues that might distinguish, for example,
an accreted dwarf galaxy from dissolved Galactic globular clusters?

\section{SEARCHING FOR STREAMS IN THE HALO}

\subsection{Historical Background}

There is a long history of searching for moving groups of stars,
also called star streams, in local samples of stars. Eggen's (1958a,b,c;
1960a,b,c,d)
work on the Hyades, Sirius, and many other possible moving groups stands out in 
particular, as well as his work on seeking moving pairs of halo
dwarfs with RR Lyraes to calibrate the luminosities of both (Eggen \&
Sandage 1959; Eggen 1977). His work
followed the ``convergent point method" for the estimation of the
distance to the Hyades. The essence of the method is that the angle
between the cluster's location and its measured convergent point provides
a relation between the radial and tangential velocities of the cluster.
Since radial velocities and proper motions are measurable, the measurement
of the angle leads to the relation between tangential velocity and
proper motion, meaning the distance. Eggen employed this technique in
assuming an association between two or more stars, determining a
convergent point, deriving distances to the individual stars, and
then testing the association by seeing if a plausible color-magnitude
diagream was the consequence. He did not, however, test for consistency
of metallicities, largely because they were not well-determined at
the time of his work.

Two other more recent methods assume the distances are relatively well
determined, and seek slightly different means to identify kinematical
relationships that point to shared dynamical histories. This is not a
simple task, for two reasons. First, stars may be relatively dispersed,
even along a common ``orbit". To counter this, recourse is often made
to ``conserved quantities" that may link stars or even
clusters travelling in somewhat 
different parts of the Galaxy's gravitational potential. Second, these
conserved quantities are few since the Galaxy's 
gravitational potential is not point-like,
and is not even very well understood. 

The first method, which we have
employed, has been used to study stars within a few hundred parsecs
of the Sun. Carney et al.\ (1996) assumed that the planar angular
momentum, $L_{\rm z}$, should be a relatively well-conserved quantity,
and in the solar neighborhood it follows 
directly from the Galactic V velocity. Total orbital
energy should also be relatively well conserved, and this may be
derived from the ``rest frame velocity", $V_{\rm RF}$:
\begin{equation}
V_{\rm RF} = U^{2} + (V + \Theta_{0})^{2} + W^{2}.
\end{equation}
Carney et al.\ (1996) suggested that structure in the V-$V_{\rm RF}$ plane
suggested that outer halo stars, now in the solar neighborhood, may have
been accreted from small galaxies. But a search for clumping of stars
in the V-$V_{\rm RF}$ plane did not reveal any ``obvious" co-moving stars
that might not have arisen by chance. By ``obvious", we mean a purely
subjective assessment of the appearance of the diagram, and ``apparent"
groupings of stars.

Helmi et al.\ (1999) employed a somewhat similar method but replaced
the orbital energy with another form of angular momentum, $L_{\phi}$, whose axis
is orthogonal to the disk's axis of rotation. They found a very intriguing
clumping of stars. Not all stars in the clump may be affiliated, however,
since the signs of their angular momenta differ, but the results
remain intriguing.

Both the above methods, and indeed any method devised to search for
moving groups, are vulnerable to the vagaries of chance. How confident
can we be that a clustering or clumping of stars in some two-dimensional,
or even three-dimensional, location is pointing to members of a dissolved
cluster or to just a random grouping? We re-investigate this
problem here.

\subsection{Searching for Streams: Our Approach}

\subsubsection{Preliminary identification of streams}

As discussed above, our sample of stars is confined to within a few
hundred parsecs of the Sun. We do not need to rely on
only two possibly conserved quantities, but we may use the full three
dimensional information contained in the U, V, and W velocities.
For a stream to maintain coherence over a period equivalent to
several Galactic orbits, stars still belonging to a stream should differ
by no more than a few kilometers per second in any of these
three velocities. This defines our basic approach. We searched through
the entire sample, looking for groups of stars that lie within a
cube of $\Delta v$ of, say, 40 \kms\ on each side in velocity space.
If a group has four or more stars, we then
derive the velocity dispersion, $\sigma$($v$), for all three velocities, requiring
that each be smaller than some small, adopted value.
Once such a group is identified, all of
its members are removed from the sample
to avoid assigning stars to multiple groups before the
search is continued.
It should be noted
that the search algorithm is not unique. Searching our sample in a
different order will find the same general groups, but perhaps with
a slightly different mix of identified members. 

\subsubsection{Monte Carlo experiments}

We must now test the significance of the possible moving groups 
found in the observed sample by
comparisons with results with those found in simulated surveys of
a model sample with
similar kinematics as ours, but which lacks any moving groups. 

We take this model sample to consist of a sphere extending 500 pc
in all directions from the observer, and with a constant star
density, an assumption that is quite reasonable for the metal-poor
halo population we're considering. The model employs $\Theta_{0}$ = 220 \kms,
and the velocity ellipsoid for the halo derived by Sommer-Larsen (1999),
with $\sigma_{\rm U} = 153$ \kms, $\sigma_{\rm V} = 93$ \kms, and
$\sigma_{\rm W} = 107$ \kms. We adopt a lower limit to proper motions
of 0.27\arcsec\ per year, typical of the Lowell Catalog which is the
primary source of our program stars. The distance limit is actually
irrelevant for the modelling because the proper motion constraint
leads to unrealistic tangential velocities for most stars at
distances of only about 300~pc. For the same reason, the
magnitude limits of the Lowell and NLTT catalogs do not affect the
results of our modelling. 

We searched for kinematical groups using the same criteria discussed
above for $\Delta v$, $\sigma$($v$), and $N_{*}$. We explored a range
of different values of $\sigma$($v$), from 40 \kms\ down to 25 \kms.
Each Monte Carlo simulation was obtained from an ensemble of $10^{4}$
simulated surveys. Table~3 compares the results from our real sample
of 692 stars with [Fe/H] $\leq$ $-1.0$ to those of the Monte Carlo
experiments. As expected, the number of possible groups identified
in the modelling rises as the $\sigma$($v$) criterion is relaxed
to larger values. So does the number in the real sample. Figure~12
compares the two sets of results graphically. {\em We conclude
that there is no compelling evidence for moving groups in our sample
of local metal-poor stars using only the three velocities as the basis
for the search.}

\begin{figure}
\epsscale{1}
\figurenum{12}
\plotone{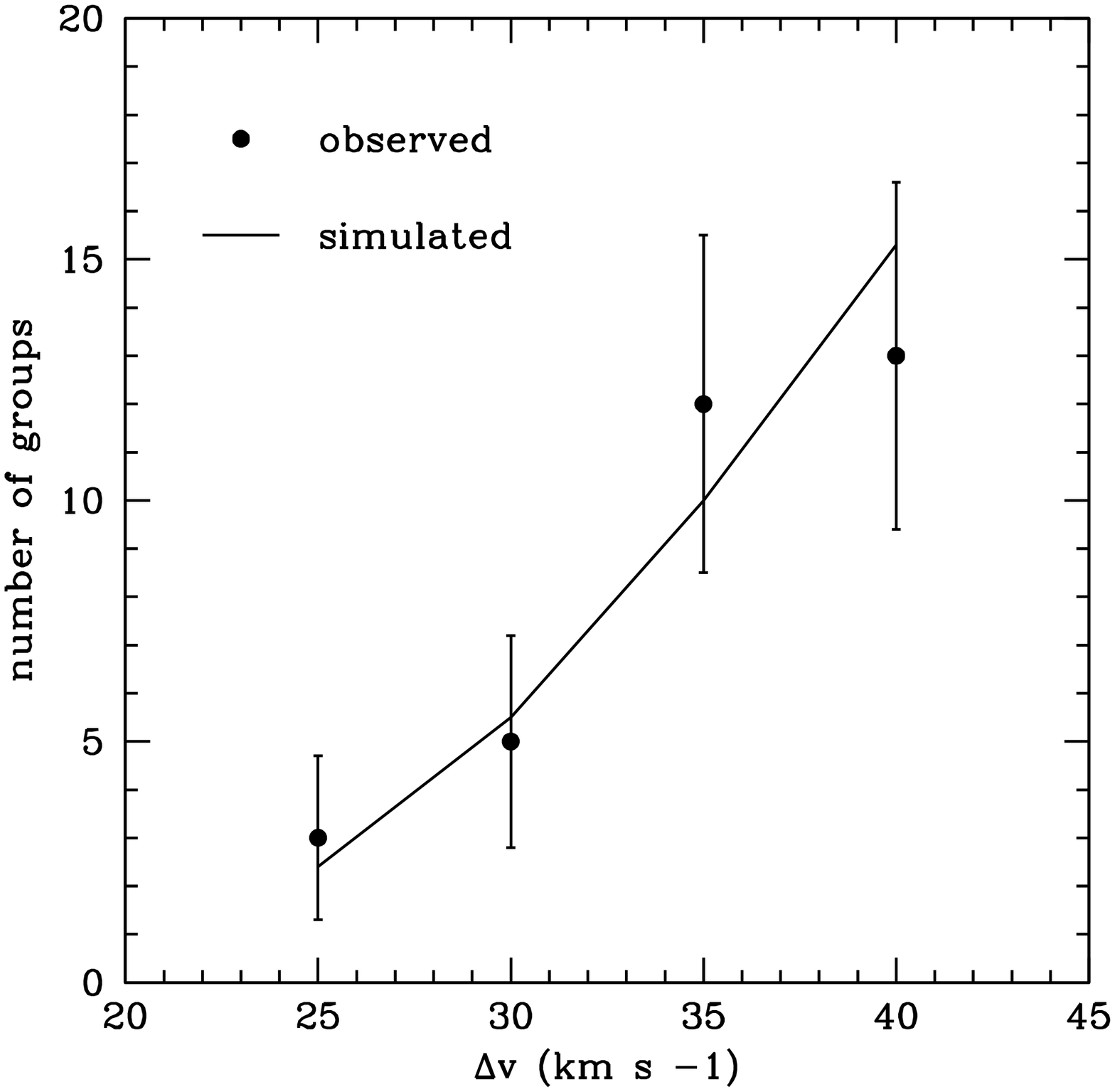}
\caption{The number of candidate kinematical groups in
our sample (dots) and from Monte Carlo experimentsi (solid
line), as a function
of $\Delta v$, the maximum velocity spread in each of the three
velocities. The error bars represent those expected using Poisson statistics.
\label{fig12}}
\end{figure}

\placefigure{fig12}

\placetable{tab3}

The above result shows that we have failed to find evidence for 
any major star stream or signs of numerous smaller star
streams with relatively small
velocity dispersionss that have an arbitrary range
of metallicities, as might be expected from accreted galaxies.

This leads, of course, to the next question: Is there evidence
for dissolved globular clusters? Here we now must include metallicity
as a search criterion, because
with the exception of $\omega$~Cen,
stars in any one globular cluster have very small
(e.g., Langer et al.\ 1998) to non-measurable differences
in [Fe/H] (Suntzeff 1993). We therefore expect 
any candidate stream to
involve stars with very similar metallicities
if it originated from a dissolved or dissolving
globular cluster.

As before, the modelling requires a good representation of the
distribution of the values of the variable, metallicity in this
case. We have adopted the measured metallicities of our program
stars, and smoothed the distribution to provide our adopted model
distribution, as shown in Figure~13.
The modelling was repeated, following exactly
that for the kinematical searches described above, but with the addition of
[Fe/H] as a variable. We varied the allowable range in the metallicity
spread, $\Delta$[Fe/H], from 0.05 up to 0.60 dex. 
(Please note that our
measured relative metallicities should be good to 0.1 dex.) Table~4
compares the results of the searches through the real sample and
the Monte Carlo simulations as a function of $\Delta v$ and
$\Delta$[Fe/H] for groups with 4 or more members. 

\begin{figure}
\epsscale{1}
\figurenum{13}
\plotone{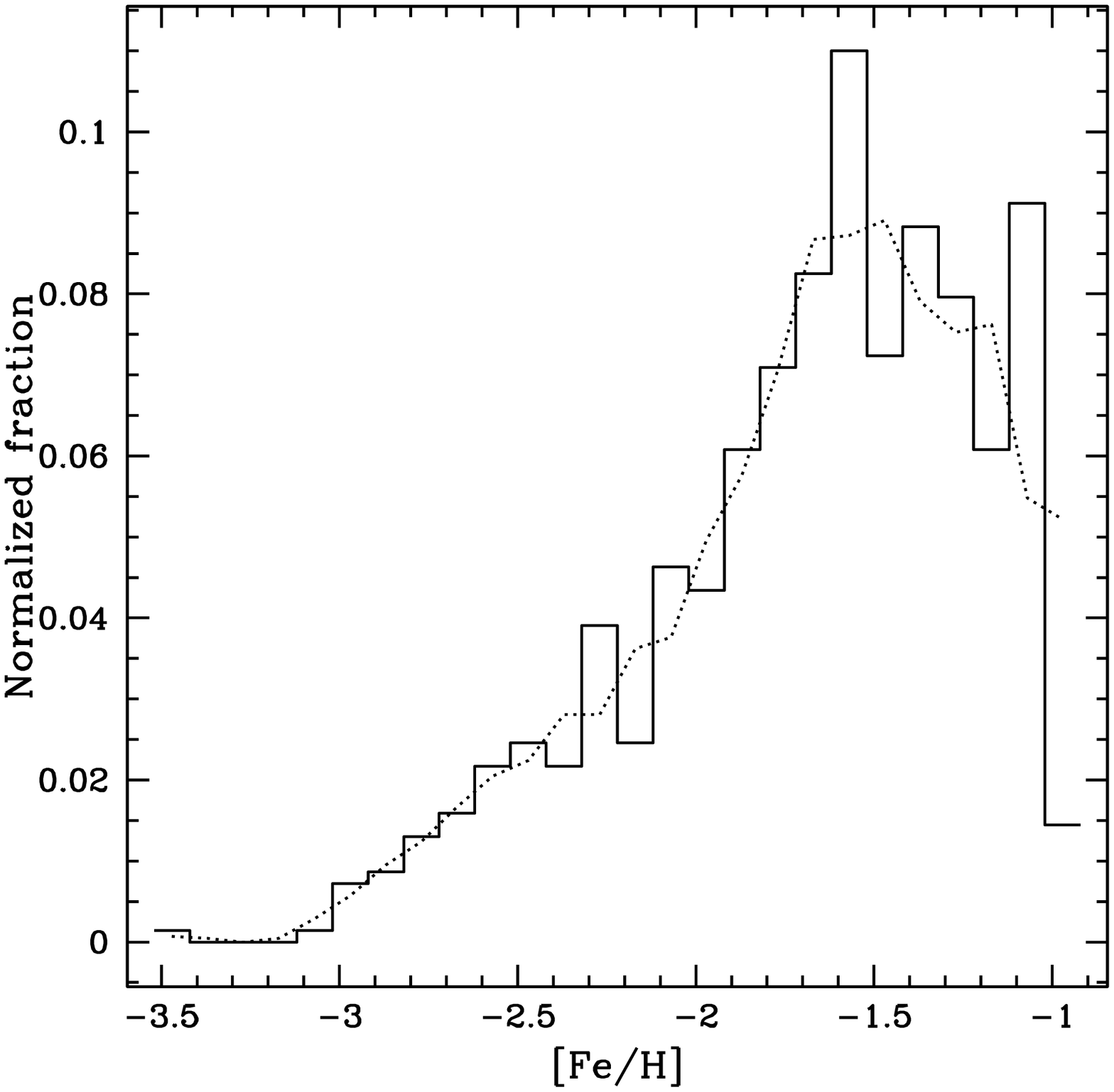}
\caption{The metallicity distributions for stars in our
program with [Fe/H] $\leq$\ $-1.0$ (solid line) and the smoothed version
used in our Monte Carlo modelling (dotted curve). \label{fig13}}
\end{figure}

\placefigure{fig13}

\placetable{tab4}

As we relax the velocity and metallicity constraints, Table~4 shows that
the number of identified moving groups in the observed sample increases,
but so does the number in the simulated surveys. Within the small
number statistics, both numbers are in fact quite similar, suggesting
that none of the identified groups are significant. There is one possible
exception. In looking in the column for $\Delta$[Fe/H] $\leq$\ 0.20,
one group in the observed sample remains, while the probability of finding
such a group with $\Delta v \leq\ 25$~\kms\ has declined to 30\%. 
The possible stream consists of
five stars, whose properties are summarized in Table~5. The group's
members have little planar orbital angular momentum, and we have
estimated their peri-Galacticon distances to lie only about 0.4 kpc
from the Galactic center. Their putative parent cluster is therefore a likely
candidate for tidal disruption during a close
passage to the Galactic center. The apo-Galacticon distance is only
about 9.6 kpc, so the cluster is unlikely to have been accreted from
another galaxy.

Even if this stream does represent a dissolved globular cluster, it falls
very far short of explaining the binary deficiency seen among
stars moving on retrograde Galactic orbits. First, while we have
noted already that tidal disruption of clusters passing near the
Galactic center could cause an effect, it would not produce
the binary deficiency seen in Figure~4. Second, the magnitude
of the effect is far too small. We have seen that to explain the
binary deficiency for stars with [Fe/H] $\leq\ -1.0$ and
V $\leq\ -220$ \kms\ requires the addition of well over one hundred
single stars. Five single stars is inadequate by a factor of at least
twenty.
We conclude that while there may be one moving group in our sample of
stars, dissolved globular clusters are unlikely to have provided a
surfeit of single stars, and hence the apparent deficiency
of binary stars, on retrograde orbits. 

\placetable{tab5}

\section{SINGLE STARS FROM ACCRETED GALAXIES?}

Globular clusters evolve dynamically, eventually populating their
outer regions with lighter single stars, and shedding them as
the evolution continues. But dwarf galaxies are hardly ideal
places to seek such segregation of single and binary stars. The
relaxation timescale is given by
\begin{equation}
\label{eq:relax}
t_{\rm relax} = 3.4 \times 10^{9} \frac{{V_{\rm m}}^{3} ({\rm km\ s^{-1}})}
{n ({\rm pc^{-3}}) [m_{*}/M_{\odot}]^{2} {\rm ln}\ \Gamma} \; {\rm years},
\end{equation}
where $n$ is the number density of stars, $V_{\rm m}$ is
the rms stellar velocity, ln~$\Gamma$ = 0.4 $\times$ N,
N is the total number of stars in the system, and
N = $M_{\rm cluster}/m_{*}$ (Spitzer 1987). Relaxation
timescales are essentially two-body processes, and the low stellar
densities in the Galaxy's current complement of satellite
galaxies appears to make dynamical relaxation timescales far too long
for the current (surviving) dwarf galaxies to have 
evolved quickly enough so that
single stars would be shed much more readily than heavier
binary stars. There are a couple of reasons, however, to not dismiss
the idea too hastily. For one,
Hurley-Keller, Mateo, and Grebel (1999) found
evidence for a central concentration of photometric binaries in
the Sculptor dwarf galaxy. Thus even though we may not understand
the single star vs.\ binary star segregation in that dwarf galaxy,
the segregation may exist nonetheless. 

We can speculate a bit about how
mergers might occur, following the discussion of Walker, Mihos,
\& Hernquist (1996).
Their work focussed on a merger involving a dwarf galaxy on a prograde
orbit on a modest inclination to the Galactic plane. The orbital
evolution and tidal dismemberment of a victim is relatively rapid,
with the consequence that the galaxy's orbit becomes nearly co-planar and
the merger is largely complete in a timescale of order 1 Gyr. They did
note, however, that the evolution is much slower for a merger involving
a dwarf galaxy on a retrograde orbit. Thus a dwarf
galaxy on a retrograde orbit has a longer time in which to
evolve, both dynamically and chemically. Walker et al.\ (1996) 
did not study this case in
as much detail, but the more specific case of the massive
globular cluster $\omega$~Centauri
has been studied by Tsuchiya, Dinescu, \& Korchagin (2003) and
Tsuchiya, Korchagin, \& Dinescu (2004). This is relevant here because
$\omega$~Cen has been suggested (Freeman 1993) to be the former nucleus
of a captured satellite galaxy, and its current Galactic orbit is
planar (W = $4 \pm 10$ \kms) and retrograde 
($\Theta = -65 \pm 10$ \kms; Dinesci, Girard, \& Van Altena 1999).
Mizutani, Chiba, \& Sakamoto (2003) suggested that the
stellar debris from such a remnant would have retrograde
motion of V $\leq\ -300$ \kms.
Unfortunately, no one has included mass segregation and other dynamical
effects in the evolution of the victim galaxy as yet. However, there is
chemical evidence that supports the idea
that $\omega$~Cen may have had a fairly long time to evolve dynamically.
Based on the enhancement of $s$-process elemental abundances,
Norris \& Da~Costa (1995),
Smith et al.\ (2000), and Rey et al.\ (2004) have suggested a chemical enrichment
history stretching over 2 to 4 Gyrs, which is consistent with the
orbital evolution timescale of Tsuchiya et al.\ (2003, 2004).

More recently, Meza et al.\ (2004; hereafter M2004) have computed
a detailed model involving the dissolution of a small
galaxy ($M \approx 4 \times 10^{9} M_{\odot}$) that leads to
a significantly increased number of stars in the Galactic halo
and a remnant moving on an orbit comparable to that of
$\omega$~Cen. As in the above simulations, the stars 
absorbed by the Galaxy are
distributed rather widely in phase space. Nonetheless, M2004
noted that there is an excess of stars in the solar neighborhood
with planar angular momenta comparable to that of $\omega$~Cen.
We are thus encouraged to seek an explanation for the
deficiency of binary stars on retrograde Galactic orbits
by seeking a possible connection with $\omega$~Cen. 

If this idea has merit, and our apparent deficiency of binary
stars on retrograde orbits is in fact caused by an excess of single
stars shed by a dwarf galaxy, 
why did it elude our detection in the search for streams? The answer
lies in the resultant velocity
dispersion of the remnant's stars in the solar neighborhood: it may be
too large for us to have detected it readily, as
the models of M2004 suggest. 

\subsection{The Chemical Footprint of $\omega$ Centauri}

Are there other ways
to detect such a system? Possibly, and here we explore the
idea that a significant fraction of the single stars moving on
retrograde orbits may be related to the putative satellite galaxy of which
$\omega$ Cen was the nucleus. We pursue this idea partly because of the
dynamical clues mentioned above, but also because the discussion will serve
as an illustrative attempt as to how to seek common origins, using
abundances as well as dynamics, since $\omega$~Cen may not,
in fact, represent the only evidence for a dissolved dwarf
galaxy moving on a retrograde Galactic orbit.

There are three separate questions to be explored. (1)~Do the field stars
have a similar mean metallicity as the former galaxy, represented here by
the globular cluster? In the absence of
knowledge about a possible relation to a surviving cluster such as
$\omega$ Cen, we can ask if the mean metallicity of the single stars on
retrograde orbits differs significantly from the 
binary stars on retrograde orbits (presumably {\em not}
contributed primarily by the accreted galaxy)
and the metal-poor single plus binary stars 
on prograde orbits. (2)~Are there any
significant differences in the detailed chemistry of these samples that
might suggest a chemical relationship with $\omega$~Cen?
In other words, do the retrograde single stars differ in [X/Fe]
patterns from the other field stars and, in this case, 
are they similar to the stars in 
$\omega$ Cen? This latter test is not definitive, 
since stars in the outer regions
of a dissolving galaxy may not have shared the same chemical evolution as
stars in the central regions represented, perhaps, by $\omega$ Cen.
(3)~If mass segregation has occured in the merging galaxy such that it shed
single stars most readily, is the central remnant, $\omega$ Cen, unusually
rich in binary stars? We explore these three questions in turn.

\subsubsection{Metallicity distribution} 

As we discussed in Section 6.2 and the introduction to this section,
a dissolved galaxy may shed stars with a wide range of metallicities,
complicating their identification in the field population. 
There are two fundamental questions to be addressed.
The simplest one is to see if the prograde and retrograde
single and binary stars show similarities or differences with the
mean metallicity of $\omega$~Cen. We
must recognize here that the cluster may not be representative
of the former galaxy's true mean metallicity or metallicity
distribution. 

We begin by defining a reference sample involving stars moving
on prograde Galactic orbits.
We have not found a significant difference
in the metallicity distributions of the single and binary stars with
[Fe/H] $\leq\ -1.0$ and moving on prograde
Galactic orbits ($-220 <$ V $\leq\ -100$ \kms), so we
merge those samples, totaling 223 stars. In Figure~14
we show the histograms of both single and
binary stars with [Fe/H] $\leq\ -1.0$ and V $\leq\ -220$ \kms.
We chose this domain on the basis of Figure~4. We show
similar results, but for
V $\leq\ -300$ \kms, in Figure~15. We chose this value
based on the
velocities of field stars from the dissolved parent galaxy of
$\omega$~Cen (Mizutani et al.\ 2003), and 
because this velocity regime shows an
even greater deficiency of binary stars (Table~1).

\begin{figure}
\epsscale{1}
\figurenum{14}
\plotone{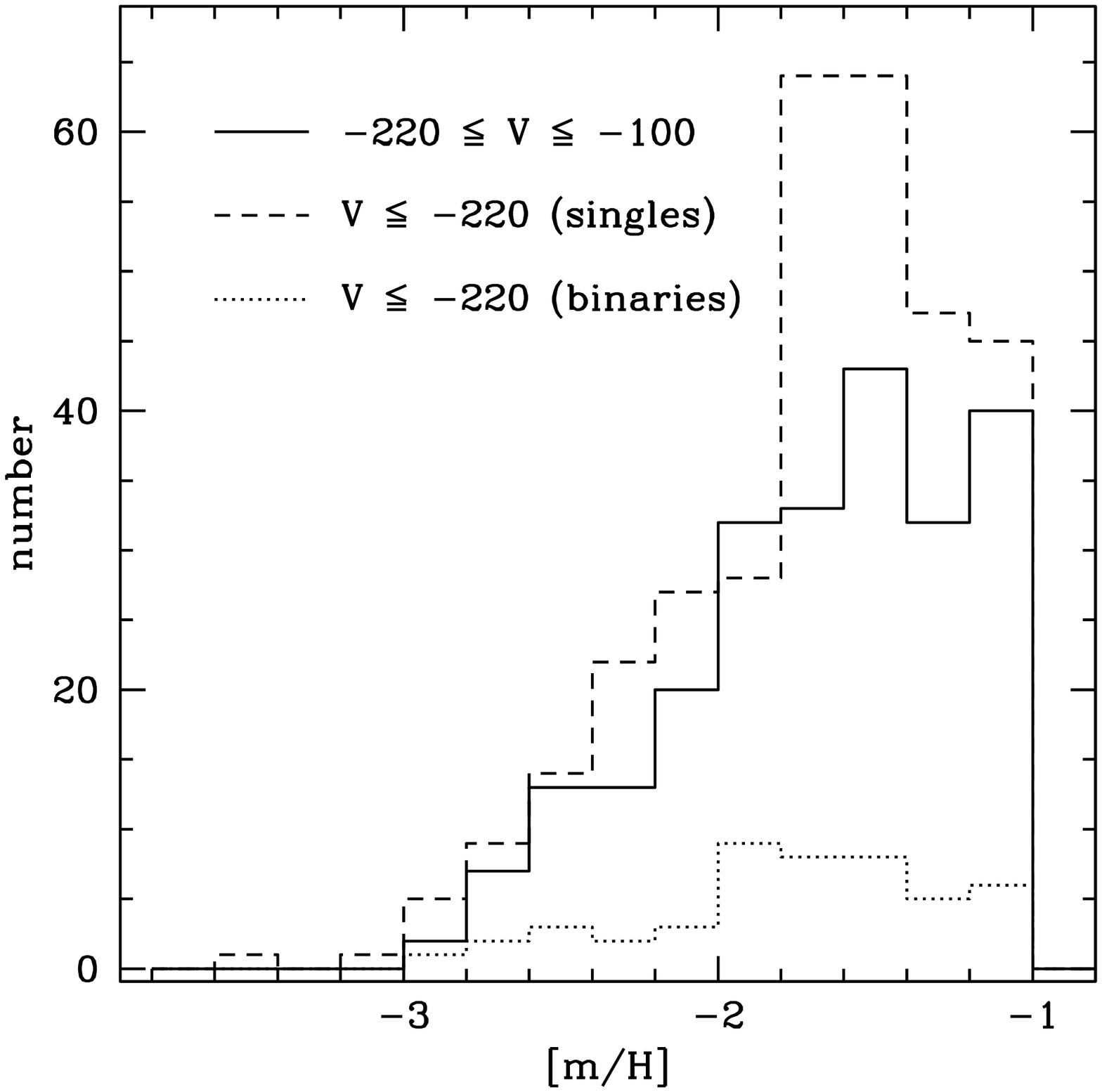}
\caption{A comparison of the distribution
of the numbers of single stars (dashed lines) and binary stars
(dotted lines) moving on retrograde orbits with V $\leq\ -220$ \kms\
compared to the total number of single+binary stars moving on prograde
orbits (solid lines), $-220 \leq$\ V $\leq\ -100$ \kms. The retrograde
condition was chosen to match the apparent domain of the binary deficiency
seen in Fig.\ 4. \label{fig14}}
\end{figure}

\placefigure{fig14}

\begin{figure}
\epsscale{1}
\figurenum{15}
\plotone{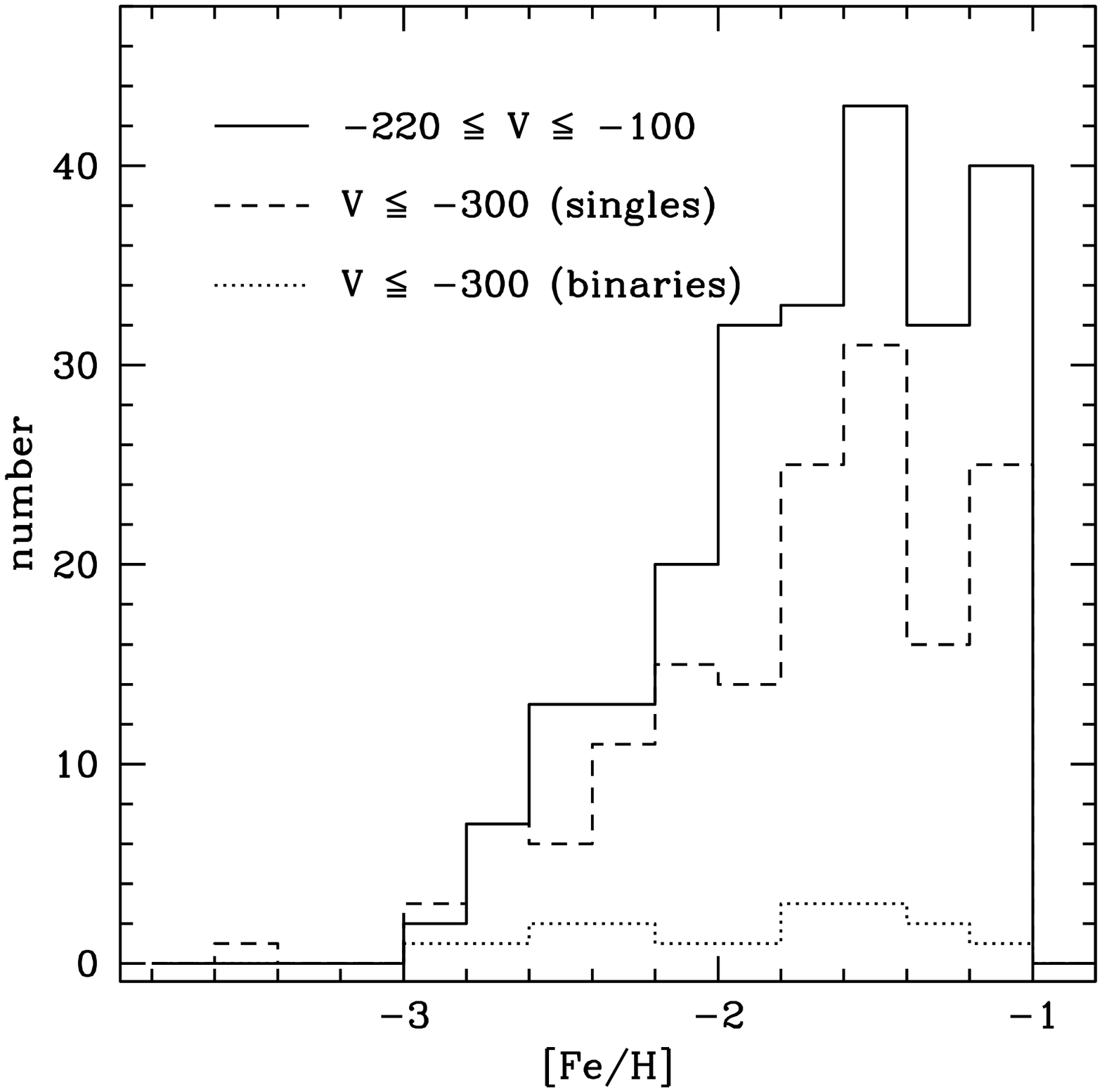}
\caption{A comparison of the distribution
of the numbers of single stars (dashed lines) and binary stars
(dotted lines) moving on retrograde orbits with V $\leq\ -300$ \kms\
compared to the total number of single+binary stars moving on prograde
orbits (solid lines), $-220 \leq$\ V $\leq\ -100$ \kms. This retrograde
condition was also chosen on the basis of Fig.\ 4, where the binary
deficiency appears to be strongest, and also from the velocities
predicted for a dissolved parent galaxy by Mizutani et al.\ (2003).
\label{fig15}}
\end{figure}

\placefigure{fig15}

The mean metallicity of $\omega$~Cen is about $-1.6$, but,
unfortunately, that is also the peak of the metallicity distribution
of the Galactic halo (consider the prograde sample; see
also Laird et al.\ 1988b; Ryan \& Norris 1991b). Figures~14 and 15
show peaks in the metallicity distributions for all the samples,
so this comparison offers us no compelling evidence for a relationship
with $\omega$~Cen. We must also expect that the stars
accreted from a dissolving dwarf galaxy will have a range
in metallicity, and, as has been known for a long time,
so do the stars of the cluster,
from [Fe/H] $\approx$ $-2.0$ to $-0.5$.

Instead of comparing the mean metallicities of field stars
to that of the cluster, we turn, instead, to testing
the predictions of our model. Specifically, do the single
stars moving on retrograde orbits, which may have arisen
in significant part from the hypothesized dissolving galaxy,
differ in their metallicity distribution from binary stars
moving on retrograde orbits, or from the single plus binary
stars moving on prograde Galactic orbits? We note that
such comparisons may not be critical tests, simply because
the Galactic halo and the accreted galaxy may well have
similar metallicity spreads, which would be expected if
both evolved as ``closed box" systems. We define the
prograde sample to be those stars with [Fe/H] $\leq\ -1.0$
and with $-220 <$ V $\leq\ -100$ \kms. Using the Kolmogorov-Smirnov
test, we find that we can reject the hypothesis that stars
in this metallicity range and with $V \leq\ -300$ \kms\ were
drawn from the same parent population as the prograde sample
with a confidence level of only 38\%. For $V \leq\ -220$ \kms,
the confidence level is 78\%, but this remains unconvincing.
The confidence levels in comparing the single 
stars and binary stars with $V \leq\ -220$ \kms\ and $V \leq\ -300$ \kms\
are not compelling either, being only 77\% and 48\%,
respectively, although the sample sizes are very small
due to the limited number of binaries moving on retrograde
Galactic orbits (47 and 17 stars, respectively). The K-S
tests do not, therefore, provide any evidence in support
of our hypothesis that a significant fraction of the
single stars moving on retrograde Galactic orbits originated
in the dissolved and accreted dwarf galaxy. (But the test
only makes sense if the metallicity spread of the galaxy
and the Galactic halo are significantly different.)

There is one interesting comparison. The prograde and retrograde
{\em binary} samples do appear to differ
in their metallicity distributions: the confidence levels
for the $V \leq\ -220$ \kms\ and $V \leq\ -300$ \kms\
samples are 90\% and 88\%, respectively. We have no
explanation for this effect.

We are therefore compelled to look more carefully
for characteristics that might distinguish $\omega$~Cen
from the normal field stars.

\subsubsection{Abundance Patterns} 

The chemical enrichment history of
$\omega$~Cen is rather different than that of other globular
clusters, as well as the field halo stars, which is one reason why
it has been suggested to be the remnant nucleus from a satellite
galaxy. Specifically, not only do its stars show a wide range
in iron abundances, but, except for the most metal-poor stars
in the cluster, its members show strong signs of enhancement
of $s$-process nucleosynthesis. Norris \& Da~Costa (1995) discussed
this characteristic at length, and their Figure~13 summarizes the situation well.
The ``light" $s$-process element Y is enhanced in the $\omega$~Cen
stars, relative to other globular clusters, and presumably single field halo
stars, by about 0.3 dex or so over the range $-1.6 \leq$\ [Fe/H] $\leq\ -1.0$.
The ``heavy" $s$-process element Ba is similarly enhanced over the same
range in [Fe/H]. At lower metallicities, the stars in $\omega$~Cen
more closely resemble other clusters and field stars in [Y/Fe] and [Ba/Fe], so
we restrict our comparisons between retrograde
and prograde field stars to this limited intermediate metallicity regime,
$-1.6 \leq$\ [Fe/H] $\leq\ -1.0$,
to maximize the differences between normal field stars and the
giants in $\omega$~Cen and still avoid trends in [Y/Fe] and [Ba/Fe]
with [Fe/H].

Do the single stars moving on retrograde Galactic orbits divide
into normal and abnormally high [Y/Fe]
or [Ba/Fe] abundance ratios?
We first re-explore the expected abundances of Y and Ba
among halo stars. The observational situation is
well reviewed by McWilliam (1997;
and references therein), and updated by Burris et al.\ (2000).
To minimize any possible systematic effects, such as might exist,
for example, between analyses of dwarf and giant stars, we
explore only the published abundance analyses of stars in our
studies of stars with large proper motions. 
To further minimize systematic effects, we employ only
two relatively large studies of the chemical abundances of field stars, those of
Fulbright (2000; hereafter F2000)
and Stephens \& Boesgaard (2002; hereafter SB2002). 
Figures~16 and 17 show
the trends of [Y/Fe] and [Ba/Fe] for single and binary stars with
[Fe/H] $\leq\ -1.0$ moving on prograde orbits. While there is a
strong decline seen in [Ba/Fe] for [Fe/H] $\leq\ -1.7$, we are
concerned only with $-1.6 \leq$\ [Fe/H] $\leq\ -1.0$, since
that is the regime in which $\omega$~Cen differs most strongly
from other clusters and from field stars. We
see that over this metallicity range, the prograde sample has
approximately constant values of [Y/Fe] and [Ba/Fe].

\begin{figure}
\epsscale{1}
\figurenum{16}
\plotone{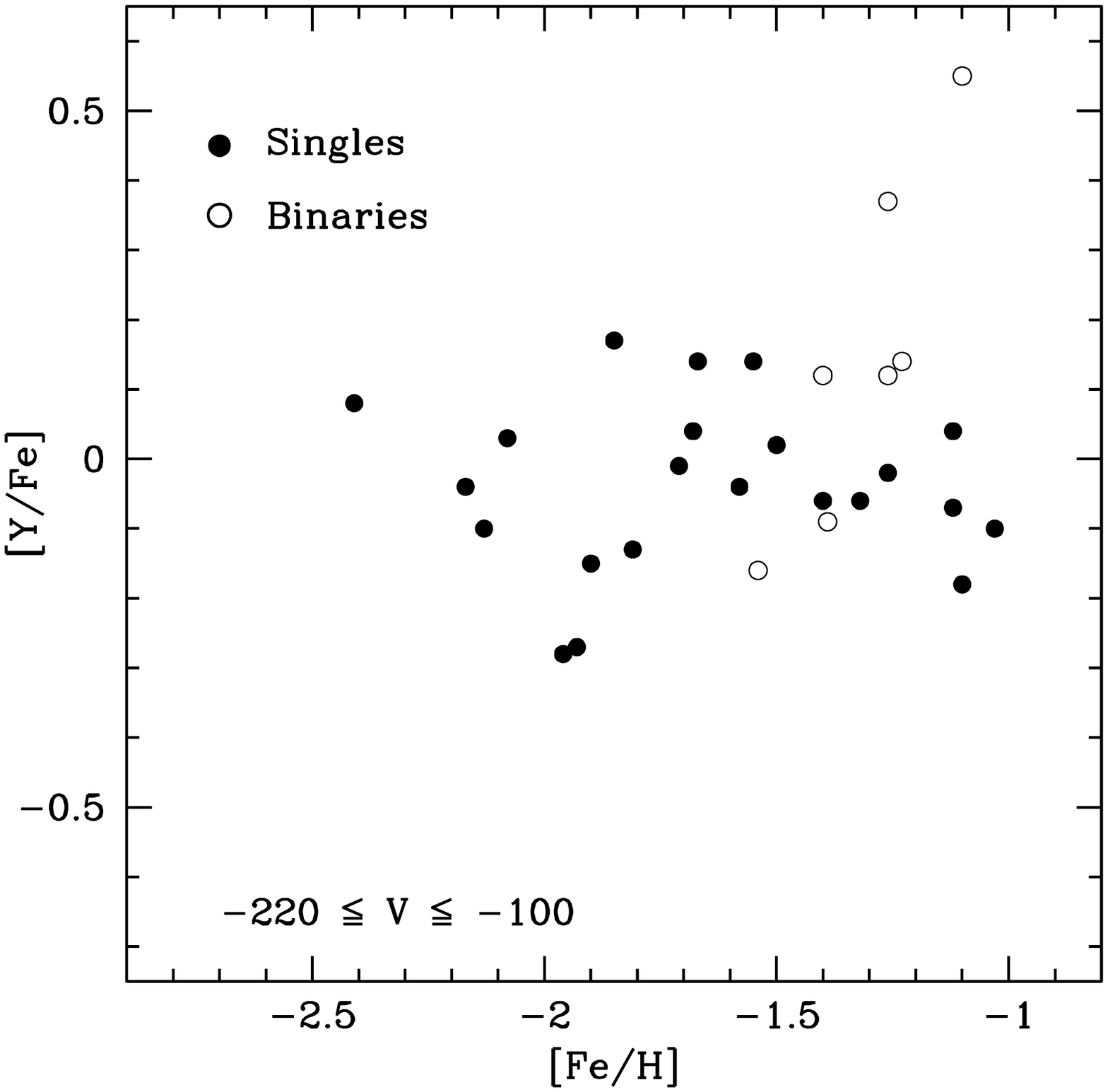}
\caption{The yttrium abundances of stars 
in our sample with data obtained from
F2000 or SB2002 moving on prograde orbits, 
$-220 \leq$\ V $\leq\ -100$ \kms, as a function of [Fe/H].
\label{fig16}}
\end{figure}

\placefigure{fig16}

\begin{figure}
\epsscale{1}
\figurenum{17}
\plotone{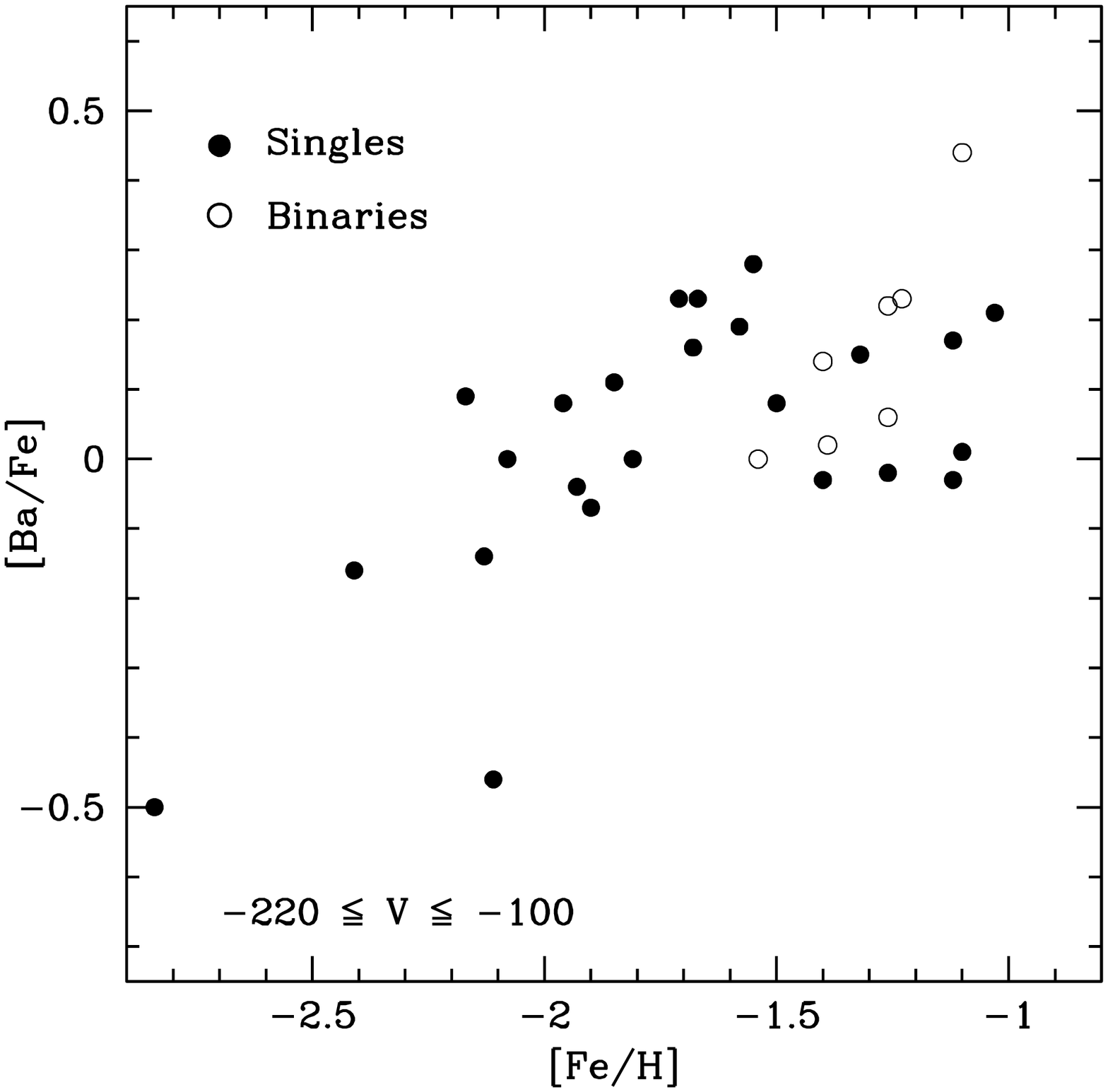}
\caption{The barium abundances of stars 
in our sample with data obtained from
F2000 or SB2002 moving on prograde orbits, 
$-220 \leq$\ V $\leq\ -100$ \kms, as a function of [Fe/H].
\label{fig17}}
\end{figure}

\placefigure{fig17}

In Figures~18 and 19 we now consider [Y/Fe] and [Ba/Fe] vs.\
the V velocity. Figure~18 hints that stars on strongly
retrograde orbits, V$ \leq\ -300$ \kms, may consist of two
groups, one with elevated [Y/Fe] values and one with 
slightly lower values of [Y/Fe]. We have provided the
basic data for these 8 single stars (and the one binary
system) in Table~6.
For the 8 single stars with
$-220 \leq$\ V $\leq\ -100$ \kms, the mean value of [Y/Fe] is
$-0.04 \pm 0.03$ ($\sigma = 0.10$), and the mean value of
[Ba/Fe] is $+0.12 \pm 0.4$ ($\sigma = 0.12$). While the mean
[Y/Fe] value for the 8 single stars with V $\leq\ -300$ \kms\
is similar to the prograde sample, $0.00 \pm 0.07$, the rms spread
is not: $\sigma = 0.21$ dex. At the risk of over-taxing a
small sample, we divide these 8 stars with V $\leq\ -300$ \kms\
further, into ``Sample A" with [Y/Fe] $>$ 0, and ``Sample B",
with [Y/Fe] $<$ 0. For A, $<$[Y/Fe]$>$ = $+0.18 \pm 0.05$
($\sigma = 0.10$), while for B, $<$[Y/Fe]$>$ = $-0.18 \pm 0.04$
($\sigma = 0.08$). The spreads are more consistent with those
obtained for the prograde sample, but the separation in [Y/Fe]
is large, $0.35 \pm 0.06$ dex. This is the same magnitude that
distinguishes the giants in $\omega$~Cen from those in other
globular clusters! Of course, dividing a sample into high
and low regions must result in a net difference between the
two groups, but let us proceed nonetheless.
Further challenging small number statistics again, it
is noteworthy that the numbers of stars in Samples A and B
are comparable, about what is needed if the low binary
fraction arises from an influx of single stars into the
retrograde sample (of order 50\%, according to the discussion
in Section~3). Qualitatively, the [Y/Fe] results
agree with the idea that the excess of single stars could
have arisen from the dissolution of the parent galaxy of
$\omega$~Cen, if it somehow had segregated single and binary stars.
However, we note that
K-S tests of the [Y/Fe] patterns for single stars on prograde
and retrograde orbits
do not confirm any significant
difference, due to the small sample sizes or 
the lack of a real difference.

\begin{figure}
\epsscale{1}
\figurenum{18}
\plotone{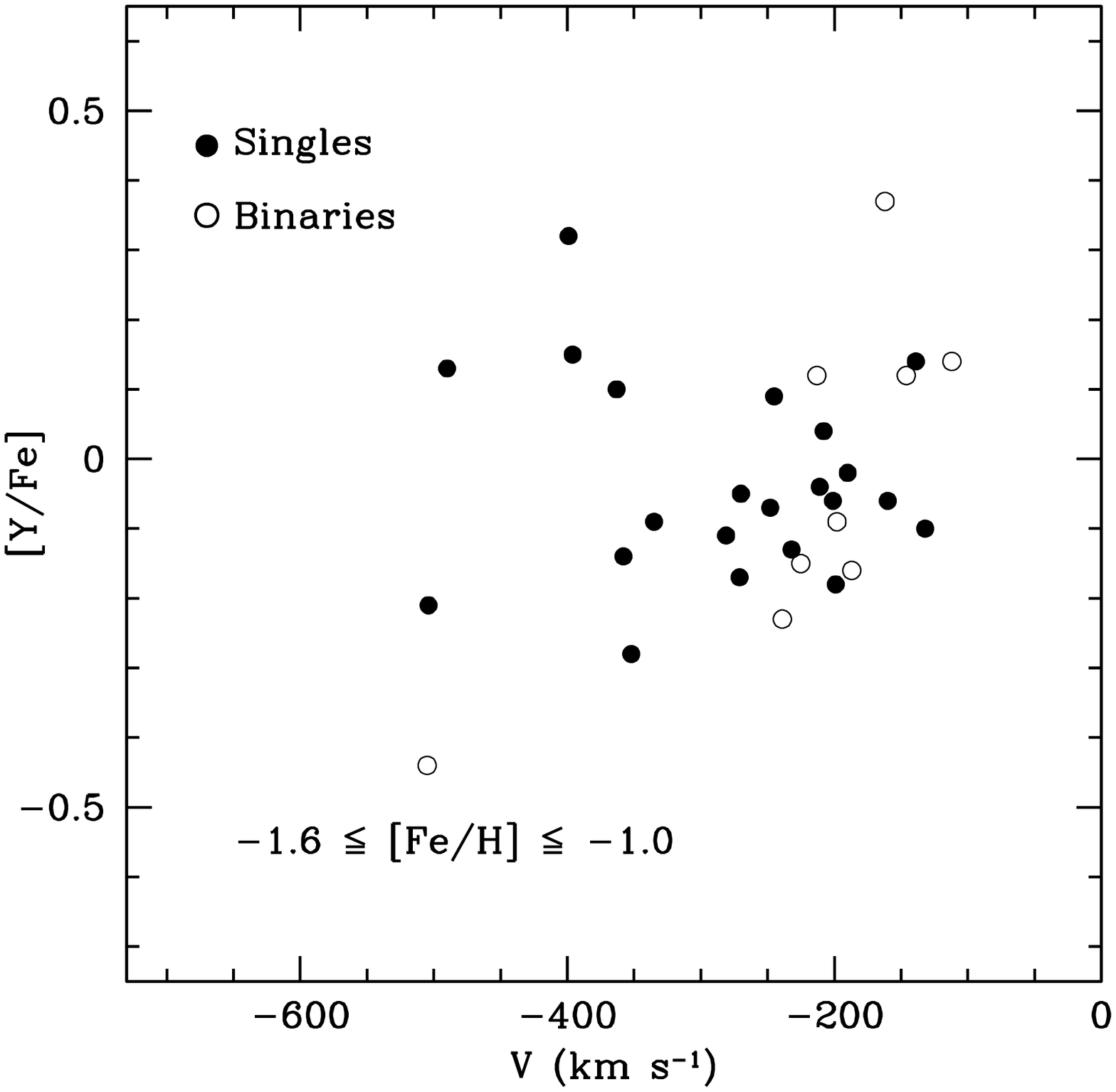}
\caption{The yttrium abundances of single and binary
stars with $-1.6 \leq$\ [Fe/H] $\leq\ -1.0$ as a function of V velocity.
\label{fig18}}
\end{figure}

\placefigure{fig18}

What about [Ba/Fe]? Figure~19 does not reveal as dramatic
a separation in [Ba/Fe] values for stars moving on retrograde
Galactic orbits.
Further, the results of Norris \& Da~Costa (1995)
suggest that the difference should be even more striking for
[Ba/Fe] than for [Y/Fe]. The K-S probabilities again do not
demonstrate any significant difference between the single
stars moving on prograde and retrograde orbits. But there
is, nonetheless a hint that differences may exist.
The 8 single stars with $-220 \leq$\ V $\leq\ -100$ \kms\ have
a mean [Ba/Fe] abundance ratio of $+0.12 \pm 0.04$ ($\sigma = 0.12$).
The 8 single stars with V $\leq\ -300$ \kms\ have a marginally lower
value, $+0.06 \pm 0.05$ ($\sigma$ = 0.13) dex. However, it is very
interesting that
we find the same four stars that define Sample A ([Y/Fe] $>$ 0) have
$<$[Ba/Fe]$>$ = $+0.16 \pm 0.04$ ($\sigma = 0.08$),
while for Sample B, $<$[Ba/Fe]$>$ = $-0.04 \pm 0.05$ ($\sigma = 0.10$).
The difference in [Ba/Fe] between Samples A and B appears to
be real, formally $0.20 \pm 0.06$ dex, but the magnitude falls short of what we expect,
$\approx 0.4$ dex,
based on the results of Norris \& Da~Costa (1995).

\begin{figure}
\epsscale{1}
\figurenum{19}
\plotone{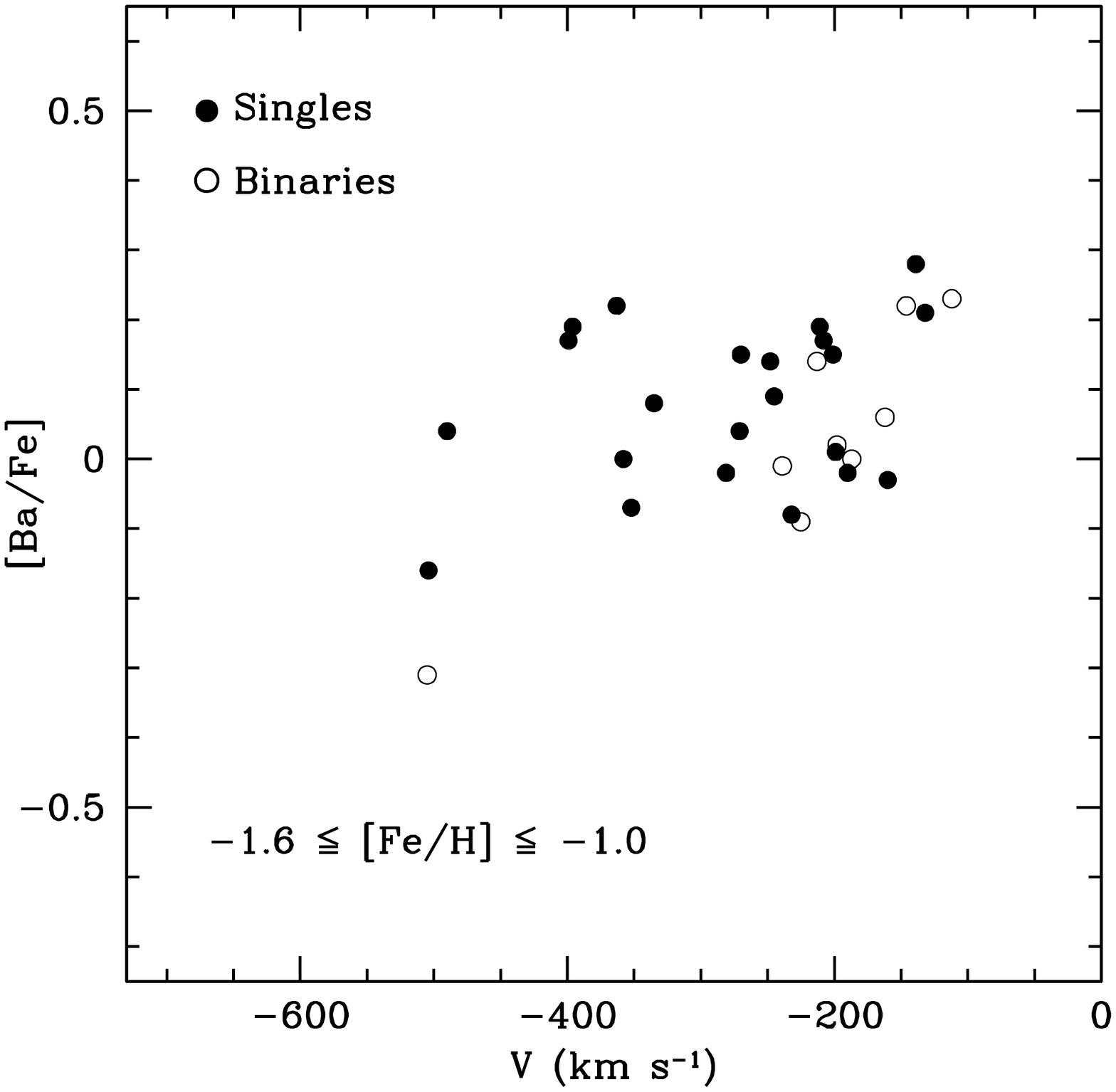}
\caption{The barium abundances of single and binary
stars with $-1.6 \leq$\ [Fe/H] $\leq\ -1.0$ as a function of V velocity.
\label{fig19}}
\end{figure}

\placefigure{fig19}

\placetable{tab6}

We conclude that the [Y/Fe] and [Ba/Fe] values hint at a bimodal
distribution in the $s$-process abundances for
stars moving on retrograde Galactic orbits.
However, the numbers of stars studied remains small, and therefore
there is not yet compelling chemical evidence 
that the single stars moving on retrograde orbits originated
from the dissolved parent galaxy of $\omega$~Cen or any
other different source. 
Larger sample sizes may prove
useful in confirming the possible differences in [Y/Fe],
and perhaps [Ba/Fe] as well. It would also be interesting to
explore the [Y/Fe], [Ba/Fe], and other $s$-process abundances with
$r$-process abundances such as Eu, in more detail. (F2000 reported
Eu abundances for only two of the single stars with retrograde
orbits, and SB2002 did not report any Eu abundances.)

M2004 have also raised the
idea that field stars that once belonged to the parent
galaxy of $\omega$~Cen may have a distinctive pattern
of the abundances of the ``$\alpha$" elements, taken here
to be Mg, Si, Ca, and Ti. M2004 employed the compilation of
abundance analyses from Gratton et al.\ (2003), and noted
that the distribution in the planar angular momenta shows
a peak at about the value expected for stars that could
be related to $\omega$~Cen or its now-dissolved parent galaxy
(Dinescu et al.\ 1999). Since the Galactic orbit of $\omega$~Cen
is now close to the Galactic plane, M2004 winnowed the Gratton
et al.\ (2003) sample further to include only 
stars with moderately retrograde
or very weakly prograde Galactic orbits, and with W velocities
($|$W$|$ $<$ 65 \kms) such that the stars do not stray
far from the plane. They found that these kinematic selection
criteria led to stars that did not show the characteristic peak
near 0 \kms\ in the distribution of U velocities, and they seized
on this to help distinguish a sample of 11 field stars (with
$|$U$|$ $>$ 50 \kms) that might be related to $\omega$~Cen. 
Those stars are indeed interesting because, like the other field
stars, they show elevated values of [$\alpha$/Fe] ($\approx +0.4$
for low metallcities, and declining [$\alpha$/Fe] values for higher
[Fe/H]). Unlike the bulk of the field stars, however, M2004 found
that this sample appears to begin the decline in [$\alpha$/Fe] at
[Fe/H] $\approx -1.5$, rather than at $-1.0$ more typical
of field halo stars. This ``earlier" decline may signal
by a slower rate of star formation, which, as we have discussed above,
appears to be a hallmark of the chemical evolution of $\omega$~Cen.

However, we do not believe that these stars are, in fact, necessarily
related to $\omega$~Cen. The primary argument against the
relationship is that Norris \& Da~Costa (1995) did not find
reduced values for any of the ``$\alpha$" elemental abundances
for cluster stars with [Fe/H] $> -1.4$. We are also concerned
by the kinematical selection criteria employed by M2004. By
considering stars with, effectively, very low V and W velocities,
stars with low U velocities will be selected against in proper motion
samples (which help define the stars analyzed by Gratton et al.\ 2003)
and do not commonly exist in this radial region of the Galaxy, where
orbital energies are typically 200 \kms. In other words,
the kinematic criteria employed by M2004 would naturally lead
to a sample with a very broad range in U velocities and an
avoidance of values near 0 \kms. This helps remove ``normal"
field stars from their sample, but we do not believe it
necessarily helps identify stars which are related to the
progenitor galaxy.

Nonetheless, the abundances of the ``$\alpha$" 
elements are interesting!
Figure~20 and Table~6 show that the stars with low $s$-process abundances
(sample B)
also show low [$\alpha$/Fe] values, with $<$[$\alpha$/Fe]$>$ = $+0.20 \pm 0.04$
($\sigma$ = 0.07 dex), while the four stars with higher $s$-process abundances
(sample A)
also show elevated [$\alpha$/Fe] abundances, with a mean of $+0.35 \pm 0.03$
($\sigma = 0.07$ dex). Further, the kinematics of the two groups of
stars are very different. 
Including the common proper motion pair (i.e., binary system)
HD~134439/134440
or not, the stars with $V \leq\ -300$ \kms\ and low $s$-process abundances
define a very hot dynamical population. Including all five stars results in
$<$U$>$ = $-85 \pm 146$ \kms\ and $<$W$>$ = $-2 \pm 57$ \kms. The velocity
dispersions ($\sigma$) are very high, 327 \kms\ for the U velocity, and
128 \kms\ for the W velocity. On the other hand, the four stars with
higher $s$-process abundances are dynamically much cooler, and perhaps
more consistent with a common origin: 
$<$U$>$ = $-57 \pm 21$ \kms\ ($\sigma =43$) \kms,
and $<$W$>$ = $+73 \pm 29$ ($\sigma = 58$) \kms. 

The kinematical differences are rather surprising, given the models
of M2004, which suggest that the dispersion in phase space for the
bulk of the dissolving galaxy should be exceptional. We show in Figure~21
the U vs.\ W velocities for the single (plus signs) and binary (circles)
stars with [Fe/H] $\leq\ -1.0$ and V $\leq\ -300$ \kms. No obvious
clustering in this limited depiction of velocity space is seen, but
neither is it expected. Perhaps the modest velocity spread found
for sample A above could be explained in the following way. If the
slow chemical enrichment seen in $\omega$~Cen was confined to the
original galaxy's central regions, then stars showing the enhanced
$s$-process abundances would have been produced only there, and
detached from the dissolving galaxy only at the end of the
accretion process. Stars detached earlier, from more the outer
regions of the original galaxy may now be dispersed more widely
in velocity space.

The referee has inquired whether we can maintain consistency
between the idea that $\omega$~Cen may have contributed the excess
of single stars on retrograde Galactic orbits, and that we
can rule out disruption of individual binary stars by
close passages to the Galactic central mass since the orbit
of $\omega$~Cen apparently comes within about one kiloparsec
of the center (Dinescu et al.\ 1999). We believe that
consideration of Figure~2 of M2004 shows
that the bulk of the disintegration of the putative massive
progenitor galaxy occurred when the perigalacticon distance
of the ensemble was larger than it is now. Hence we would
argue that our results would still rule out the disruption of
individual binary systems because the binary deficiency
extends to stars with large (but negative) V velocities,
but the dissolution of the galaxy would have produced stars
with rather different kinematics than is the case
now for $\omega$~Cen. What we probably must say, however,
is that we cannot invoke evaporation of single and binary
stars from $\omega$~Cen, and that the binaries are
subsequently disrupted due to the close passage to
the Galactic central mass. 

\subsubsection{Binary Fraction} 

Mayor et al.\ (1997) summarized radial velocity measurements for
almost five hundred bright red giants in $\omega$~Cen. While, so
far as we are aware, no analysis has been published as yet regarding
the binary characteristics of this sample, their Table~1 provides
enough information to obtain a preliminary estimate of the frequency
of binary stars in the cluster.
Specifically, the $\chi^{2}$
probability that the velocity variations could have arisen via
observational uncertainties is a very useful tool. In the absence
of any real velocity variations, P($\chi^{2}$) should be a flat
distribution, but as Figure~22 shows, there is a very large spike
at the smallest values. While only 5\% of the stars should have
probability values in the range between 0.00 and 0.05, roughly
one third of the stars have enough radial velocity
variability that their P($\chi^{2}$) have
such low values. Taken at face value,
this appears to confirm a very high frequency of binary stars among
the bright red giants in the cluster, well in excess of the
binary frequency found by Carney et al.\ (2003) for metal-poor
field red giants ($\approx 17$\%). However, as Gunn \& Griffin (1979)
were the first to note, luminous metal-poor red giants in the globular
cluster M3 show radial velocity variability that is not obviously
periodic and due to orbital motion. 
This apparently aperiodic velocity variation has become known as ``velocity jitter". 
Roughly 40\% of field
red giants more luminous than $M_{V} \approx -1.4$ show jitter
(Carney et al.\ 2003), as do luminous red giants in several
other globular clusters (Mayor \& Mermilliod 1984; Lupton,
Gunn, \& Griffin 1987; Pryor, Hazen, \& Latham 1988;
C\^{o}t\'{e} et al.\ 1996). In fact, if we restrict the
$\omega$~Cen sample to only those stars with $V > 12.3$ ($M_{V} > -1.4$),
Figure~20 shows that  spike at the lowest probability 
values declines significantly,
and the evidence for binaries among the red giants in $\omega$~Cen
is weakened considerably. The spike is now only 23\% of the sample,
compared to the expected 5\%. Thus the ``excess" of $\approx 18$\%
is consistent with a binary fraction expected from the
metal-poor field red giants. However, the number of velocities
for each star is generally small, so that the excess of $\approx 18$\% is
certainly a lower limit. We need to know how many binaries could have
eluded detection with such a data set, before we compare the
binary fraction in $\omega$~Cen with the metal-poor field stars.
And, of course, additional observations
of the giants in $\omega$~Cen would be highly desirable.

\placefigure{fig21}

\section{SUMMARY}

A probable deficiency of binaries has been found among metal-poor
stars whose V velocities indicate retrograde rotation. Deficiencies
are not seen at extreme U or W velocities. Theory enables us
to rule out preferential disruption of field binary stars on high-velocity
retrograde orbits.
The dominance of the binary deficiency
in a single velocity direction rules out ejection of single
stars from globular clusters.
Stars with low angular momentum,
whose orbits carry them close to the Galactic center, do not
show such a deficiency of binaries, ruling out tidal effects
from the Galactic central mass concentration. We have explored
carefully the idea that dynamically evolved stellar systems,
which might shed single stars most readily, could explain the
binary deficiency. A careful study of our sample and Monte Carlo
modelling reveals that perhaps only one dissolved globular
cluster has contributed to our sample, and only a few stars. It is
of interest that this stream passes near the
Galactic center, where strong tidal effects may have dissolved
the parent cluster.
We also investigated the possibility that a single dwarf galaxy moving
on a roughly planar retrograde orbit could have caused the
effect, due to some as-yet unidentified mechanism that would
have segregated its lower mass single stars to its outer
regions while concentrating it higher mass binary stars more
centrally. Specifically, we explored the possibility that
a significant fraction of the single stars on retrograde orbits
were detached from the outer regions of the parent galaxy
of the globular cluster $\omega$~Cen through a comparison
of mean metallicities; [Y/Fe], [Ba/Fe], and [$\alpha$/Fe];
and binary fractions.
The chemical abundances provided some support to the idea,
but additional spectroscopic studies are needed.

\acknowledgments

BWC is grateful to the National Science Foundation for grant
support to the University of North Carolina (AST-9619831,
AST-9988156, and AST-030305431). JBL similarly thanks the NSF
for support to Bowling Green State University for grants
AST-9619628, AST-9988247, and AST-0307340. LA thanks DGAPA/UNAM
for support through grant IN113403. We thank the anonymous
referee for a careful review and some thought-provoking
suggestions that have helped improve the manuscript.

\section*{Appendix A: Binary Disruption due to encounters with black holes}

The energy transfer from a perturbing mass to a binary system is a complicated
phenomenon. For example, it is difficult to take into account
the relative motions of the binary
components while the interaction takes place. In the limit where this motion
can be approximated as a 1--dimensional harmonic oscillator, it is possible to
compute the velocity ``kick'' produced by the tidal force of a point
mass perturber as (Spitzer 1958):
$$
\Delta {\bf v} = \left({{2GM_{\rm p}}\over{b^2v_{\rm p}}}\right)\, L(\beta) \,{\bf b},
\eqno(A1)
$$
where $v_{\rm p}$ is the velocity of the perturbing mass $M_{\rm p}$, ${\bf b}$ is the impact
parameter vector whose magnitude is given by the minimum distance between perturbed
and perturbing masses and its direction points from the former to the latter.
$L$ is a correction for adiabaticity and it is a complicated function that must
be computed numerically. We will use the following analytical approximation:
$$
L(\beta) = \cases{ 1, & if $\beta\le \beta_o= 1.64$;\cr
                   e^{-\lambda(\beta-\beta_o)}, & if $\beta>\beta_o$;\cr}
\eqno(A2)
$$
where $\lambda=1.082$ and $\beta$ is given by the ratio:
$$
\beta\equiv \omega_* \left({{2b}\over{v_{\rm p}}}\right).
\eqno(A3)
$$
Here $\omega_*$ is the angular frequency of the binary system. This ratio corresponds
to $2\pi$ times the ratio of the encounter time to binary period that we used in
section~4.2 as a condition for impulsiveness.

For our computations, we express $\beta$ in terms of the binary 
separation $a$,
using the circular orbit approximation, $\omega_*^2 = (2Gm_*)/a^3$:
$$
\beta = 2\sqrt{2G} \left({m_*\over a^3}\right)^{1/2} \left({b\over v_{\rm p}}\right)
\eqno(A4)
$$
In physical units, this can be written as
$$
\beta = 84.249\,
        \left[{{(m_*/M_\odot)}\over{(a/AU)^3}}\right]^{1/2}\,
        \left[{{(b/AU)}\over{v_{\rm p}/(\rm km~s^{-1})}}\right]
\eqno(A5)
$$
The change in energy per unit mass corresponding to this velocity perturbation is
$$
\Delta E = \Delta\left({{1\over 2}\,|{\bf v}|^2}\right)
         = ({\bf v}\cdot\Delta{\bf v}) + (1/2)|\Delta{\bf v}|^2
\eqno(A6)
$$
The first (linear) term in this expression depends on 
the relative orientation of
${\bf v}$ and the perturber at closest approach. The magnitude 
of the effect can be
up to $|v\Delta v|$ (when the stars in the binary are moving 
along the line defined by
the impact parameter), but its average value is zero. The 
second (quadratic) term is
always positive and thus disruptive to the binary. Since the 
binary members are most
likely to move quite a bit during the interaction, 
the linear term
will average to zero, and so we will use only the second term as an
estimate for average heating:
$$
\Delta E = 2\left[{{GM_p}\over{bv_{\rm p}}}\,L(\beta)\right]^2.
\eqno(A7)
$$

We now compare this energy change with the binding energy 
(per unit mass) of the binary,
$E_b = -Gm_*/2a$:

$$
{{(\Delta E)}\over{|E_b|}} = 4G \left({a\over m_*}\right)
                                    \left({M_{\rm p}\over{bv_{\rm p}}}\,L(\beta)\right)^2.
\eqno(A8)
$$
In terms of physical units this can be written as
$$
{{(\Delta E)}\over{|E_{\rm b}|}} = 3.549\times 10^3
                         \left[{{(a/AU)}\over {(m_*/M_\odot})}\right]
                         \left[{{(M_{\rm p}/M_\odot)}\over{(b/AU)(v_{\rm p}/({\rm km~s^{-1}))}}}
			 \,L(\beta)\right]^2.
\eqno(A9)
$$

In Figure~A1 we plot the location of the $(\Delta E/|E_b|) = 1$ 
line in the binary
separation vs.\ impact parameter plane. Two boundaries are 
plotted for encounter
velocities of $10$ and $400$~\kms. Points below and 
to the right of the boundaries
correspond to strong encounters that destroy the binary. 
Even for a collision with the higher relative velocity, these
values imply that very close encounters were involved. (We should
recall that $a$ is less than a few AU for most of the binaries
in our sample, which implies that $b$ is generally less than
$\approx$ 0.01 pc.)
A population
of massive black holes ($10^6\, M_\odot$ in our calculation) 
with a number density
large enough to produce close encounters like those required to destroy the
binaries in our sample, would have a huge dynamical effect on
Galactic stellar motions that has not been seen.
Note also that throughout the calculations in this Appendix we have
assumed a point particle perturber, an extreme situation. Extended
perturbers such as giant molecular clouds would have even weaker
disruption effects.

\placefigure{figa1}

\clearpage

\begin{deluxetable}{lrrrrrrrrr}
\footnotesize
\tablewidth{0pc}
\tablenum{1}
\tablecaption{Binary Fractions vs. Kinematics \label{tab1}}
\tablehead{\colhead{} & \colhead{} & \colhead{MAIN} & \colhead{} &
 \colhead{} & \colhead{RYAN} & \colhead{} & \colhead{} &
 \colhead{COMB} & \colhead{} \\
\\
\colhead{Range} &
\colhead{N(S)} &
\colhead{N(B)} &
\colhead{$f$} &
\colhead{N(S)} &
\colhead{N(B)} &
\colhead{$f$} &
\colhead{N(S)} &
\colhead{N(B)} &
\colhead{$f$}}
\startdata
\multicolumn{10}{c}{}\nl
\multicolumn{10}{c}{(a) Metallicity}\nl
\multicolumn{10}{c}{}\nl
[Fe/H] $\leq$\ $-1.0$ & 326 & 84 & $21 \pm 2$\% & 189 & 33 & $15 \pm 3$\% &
515 & 117 & $19 \pm 2$\% \nl
[Fe/H] $>$ $-1.0$ & 430 & 154 & $26 \pm 2$\% & 160 & 30 & $16 \pm 3$\% &
590 & 184 & $24 \pm 2$\% \nl
\multicolumn{10}{c}{}\nl
\multicolumn{10}{c}{(a) Metallicity; V $> -220$}\nl
\multicolumn{10}{c}{}\nl
[Fe/H] $\leq$\ $-1.0$ & 121 & 56 & $31 \pm 4$\% & 61 & 16 & $21 \pm 5$\% &
182 & 72 & $28 \pm 3$\% \nl
[Fe/H] $>$ $-1.0$ & 401 & 154 & $28 \pm 2$\% & 126 & 28 & $18 \pm 3$\% &
527 & 182 & $26 \pm 2$\% \nl
\multicolumn{10}{c}{}\nl
\multicolumn{10}{c}{(b) V velocity; [Fe/H] $\leq$\ $-1.0$} \nl
\multicolumn{10}{c}{}\nl
V $>$ $-220$ & 121 & 56 & $32 \pm 4$\% & 61 & 16 & $21 \pm 5$\% &
182 & 72 & $28 \pm 3$\% \nl
V $\leq$\ $-220$ & 205 & 30 & $13 \pm 2$\% & 122 & 17 & $12 \pm 3$\% &
327 & 47 & $13 \pm 2$\% \nl
V $\leq$\ $-300$ & 95 & 9 & $9 \pm 3$\% & 58 & 8 & $12 \pm 4$\% &
153 & 17 & $10 \pm 2$\% \nl
\multicolumn{10}{c}{}\nl
\multicolumn{10}{c}{(c) $|$U$|$ velocity; [Fe/H] $\leq$\ $-1.0$} \nl
\multicolumn{10}{c}{}\nl
$|$U$|$ $\geq$ 200 & 135 & 32 & $19 \pm 3$\% & 64 & 12 & $16 \pm 5$\% &
199 & 44 & $18 \pm 3$\% \nl
$|$U$|$ $<$ 200 & 191 & 52 & $21 \pm 3$\% & 125 & 21 & $14 \pm 3$\% &
316 & 73 & $19 \pm 2$\% \nl
\multicolumn{10}{c}{}\nl
\multicolumn{10}{c}{(d) $|$W$|$ velocity; [Fe/H] $\leq$\ $-1.0$} \nl
\multicolumn{10}{c}{}\nl
$|$W$|$ $\geq$ 80 & 127 & 32 & $20 \pm 4$\% & 82 & 9 & $10 \pm 3$\% &
209 & 41 & $16 \pm 3$\% \nl
$|$W$|$ $<$ 80 & 199 & 52 & $21 \pm 3$\% & 107 & 24 & $18 \pm 4$\% &
306 & 76 & $20 \pm 2$\% \nl
\multicolumn{10}{c}{}\nl
\multicolumn{10}{c}{(e) W velocity; [Fe/H] $\leq$\ $-1.0$} \nl
\multicolumn{10}{c}{}\nl
W $\geq$\ 100 & 49 & 7 & $13 \pm 5$\% & 18 & 1 & $5 \pm 5$\% &
67 & 8 & $11 \pm 4$\% \nl
\tablevspace{4pt}
\enddata
\end{deluxetable}

\clearpage

\begin{deluxetable}{lrrrr}
\footnotesize
\tablewidth{0pc}
\tablenum{2}
\tablecaption{Field and M13 Luminosity Functions \label{tab2}}
\tablehead{
\colhead{M13 $V$ mag} & \colhead{($B-V$)$_{0}$} & 
\colhead{Field} & \colhead{M13} &
\colhead{Normalized Field}} 
\startdata
18-19 & 0.400 - 0.414 & 2 & 20 & 0.9 \nl
19-20 & 0.414 - 0.531 & 94 & 43.3 & 43.3 \nl
20-21 & 0.531 - 0.713 & 87 & 81.0 & 40.1 \nl
21-22 & 0.713 - 0.945 & 42 & 94.3 & 19.3 \nl
 $>$ 22 & $>$ 0.945 & 0 & 735.0 & 0 \nl
\tablevspace{4pt}
\enddata
\end{deluxetable}

\clearpage

\begin{deluxetable}{lcc}
\footnotesize
\tablewidth{0pc}
\tablenum{3}
\tablecaption{Comparisons of Numbers of Kinematical Groups \label{tab3}}
\tablehead{
\colhead{$\sigma$($v$) (km sec$^{-1}$)} & \colhead{Observed sample} & 
\colhead{Simulated sample}} 
\startdata
\tablevspace{4pt}
40 & $13 \pm 3.6$ & 15.3 \nl
35 & $12 \pm 3.5$ & 10.0 \nl
30 & $5 \pm 1.7$ & 5.5 \nl
25 & $ 3 \pm 1.7$ & 2.4 \nl
\tablevspace{4pt}
\enddata
\end{deluxetable}

\clearpage

\begin{deluxetable}{lrrrrrrrr}
\footnotesize
\tablewidth{0pc}
\tablenum{4}
\tablecaption{Searches for Metallicity-based Moving Groups; $N_{*} \geq\ 4$ \label{tab4}}
\tablehead{
\colhead{$\Delta v$} & \colhead{$\Delta$[Fe/H] $<$\ 0.05} & \colhead{0.10} &
\colhead{0.15} & \colhead{0.20} & \colhead{0.30} & \colhead{0.40} &
\colhead{0.50} & \colhead{0.60} }
\startdata
\tablevspace{4pt}
$\leq$\ 40 & 0 (0.12) & 0 (0.71) & 2 (1.6) & 2 (2.6) & 3 (4.1) & 5 (5.1) & 6 (5.9) & 10 (6.5) \nl
$\leq$\ 35 & 0 (0.08) & 0 (0.43) & 0 (1.0) & 1 (1.6) & 2 (2.6) & 4 (3.2) & 4 (3.8) & 8 (4.2) \nl
$\leq$\ 30 & 0 (0.03) & 0 (0.20) & 0 (0.5) & 1 (0.8) & 1 (1.2) & 1 (1.5) & 1 (1.8) & 2 (2.1) \nl
$\leq$\ 25 & 0 (0.01) & 0 (0.08) & 0 (0.2) & 1 (0.3) & 1 (0.5) & 1 (0.6) & 1 (0.8) & 2 (0.9) \nl
\tablevspace{4pt}
\tablecomments{Main entries in the table are the numbers of groups found in the program sample.
The numbers in parentheses are the average number found in $10^{4}$ Monte Carlo simulations
of a sample of equal size, as described in the text.}
\enddata
\end{deluxetable}

\clearpage

\begin{deluxetable}{lrrrrrr}
\footnotesize
\tablewidth{0pc}
\tablenum{5}
\tablecaption{A Possible Star Stream \label{tab5}}
\tablehead{
\colhead{Star} & \colhead{[Fe/H]} & 
\colhead{U} & \colhead{V} & \colhead{W} &
\colhead{R$_{\rm apo}$} & \colhead{R$_{\rm peri}$}} 
\startdata
       &  &  &  &  &  &  \nl
HD~3567 & $-1.36$ & $-144$ & $-201$ & $-28$ & 9.7 & 0.3 \nl
LP~646-4 & $-1.25$ & $-136$ & $-191$ & $-29$ & 9.7 & 0.5 \nl
LP~815-7 & $-1.40$ & $-154$ & $-223$ & $-65$ & 9.7 & 0.1 \nl
BD$-10$~5549 & $-1.26$ & $-128$ & $-184$ & $-45$ & 9.5 & 0.6 \nl
LFT~1697 & $-1.43$ & $-135$ & $-195$ & $-36$ & 9.6 & 0.4 \nl
       &  &  &  &  &  &  \nl
mean & $-1.34$ & $-139$ & $-199$ & $-41$ & 9.6 & 0.4 \nl
$\sigma$ & $\pm 0.08$ & $\pm 10$ & $\pm 15$ & $\pm 15$ & $\pm 0.1$ & 
$\pm 0.2$ \nl
\tablevspace{4pt}
\enddata
\end{deluxetable}

\begin{deluxetable}{lrrrrrrrr}
\footnotesize
\tablewidth{0pc}
\tablenum{6}
\tablecaption{Abundance Differences for Stars with V $\leq\ -300$ \kms \label{tab6}}
\tablehead{
\colhead{Star} & \colhead{class} & \colhead{[Fe/H]} & 
\colhead{[Y/Fe]} & \colhead{[Ba/Fe]} & \colhead{[$\alpha$/Fe]} &
\colhead{U} & \colhead{V} & \colhead{W} }
\startdata
G189-50 & single & $-1.47$ & $-0.28$ & $-0.07$ & +0.12 & +294 & $-352$ & $-109$ \nl
G93-1 & single & $-1.59$ & $-0.21$ & $-0.16$ & +0.26 & $-375$ & $-504$ & $-107$ \nl
G241-4 & single & $-1.57$ & $-0.14$ & +0.00 & +0.15 & $-280$ & $-358$ & +168 \nl
G31-26 & single & $-1.32$ & $-0.09$ & +0.08 & +0.25 & +249 & $-335$ & +101 \nl
HD 134439/40 & binary & $-1.54$ & $-0.21$ & $-0.16$ & +0.15 & $-312$ & $-505$ & $-64$ \nl
 & & & & & & & & \nl
HD 132475 & single & $-1.59$ & +0.10 & +0.22 & +0.44 & $-43$ & $-363$ & +70 \nl
G97-40 & single & $-1.52$ & +0.13 & +0.04 & +0.28 & $-118$ & $-490$ & +75 \nl
G170-21 & single & $-1.47$ & +0.15 & +0.19 & +0.33 & $-48$ & $-396$ & +2 \nl
G46-5 & single & $-1.41$ & +0.32 & +0.17 & +0.33 & $-19$ & $-399$ & +144 \nl
\enddata
\end{deluxetable}

\clearpage


\clearpage

\clearpage

\clearpage

\clearpage

\clearpage

\clearpage

\clearpage

\clearpage

\clearpage

\clearpage

\clearpage

\clearpage

\clearpage

\clearpage

\clearpage

\clearpage

\clearpage

\clearpage

\clearpage

\clearpage

\begin{figure}
\epsscale{1}
\figurenum{20}
\plotone{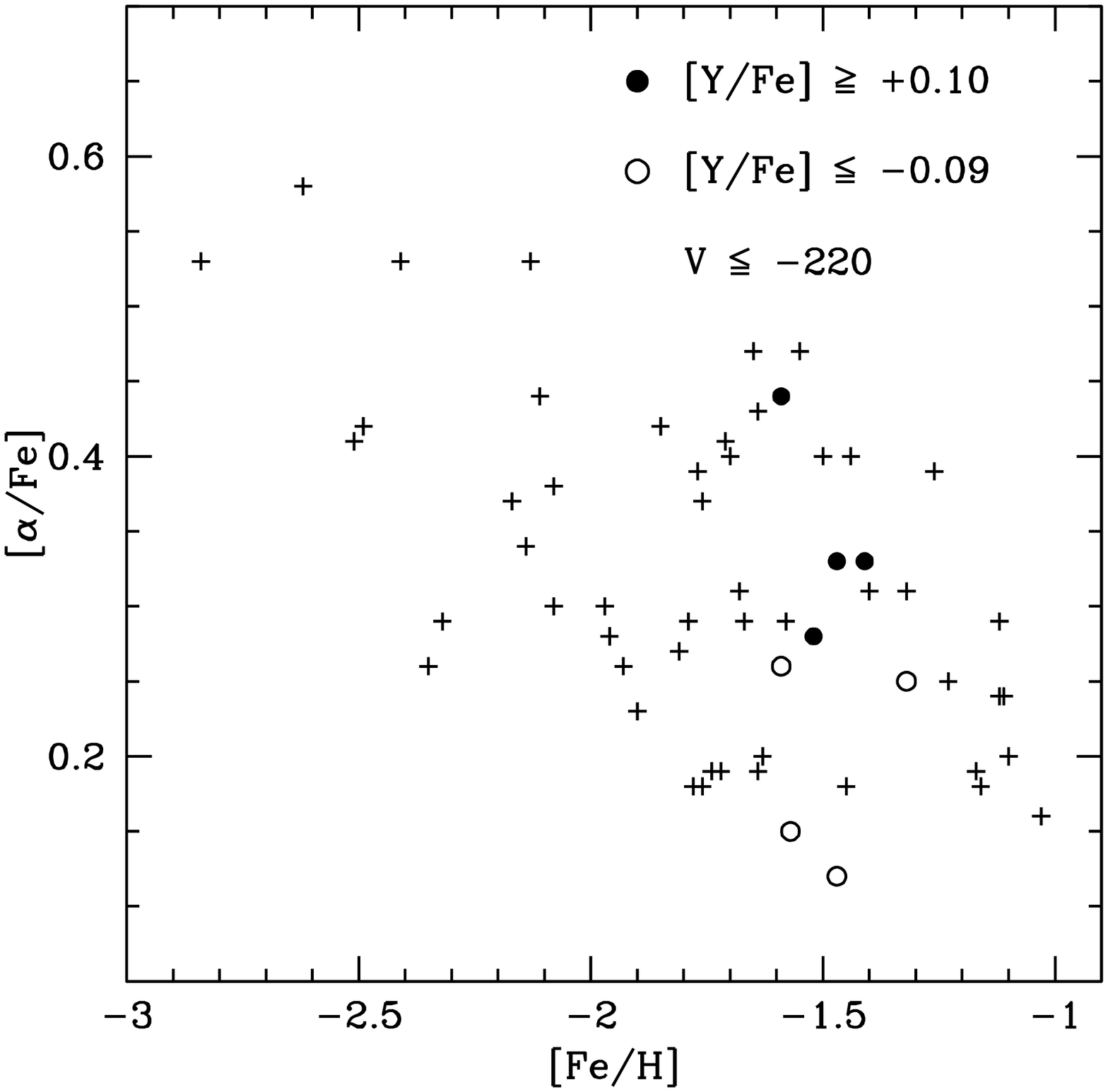}
\caption{The behavior of [$\alpha$/Fe] vs.\ [Fe/H] for stars in
our program, with data reported by F2000 or SB2002.
The four stars with V $\leq$\ $-300$ \kms and
higher [Y/Fe] and [Ba/Fe] abundances shown
in Figures~18 and 19, and discussed in the text, are identified
by filled circles. The four stars with lower [Y/Fe] and [Ba/Fe]
values are shown as open circles.
\label{fig20}}
\end{figure}

\clearpage

\begin{figure}
\epsscale{1}
\figurenum{21}
\plotone{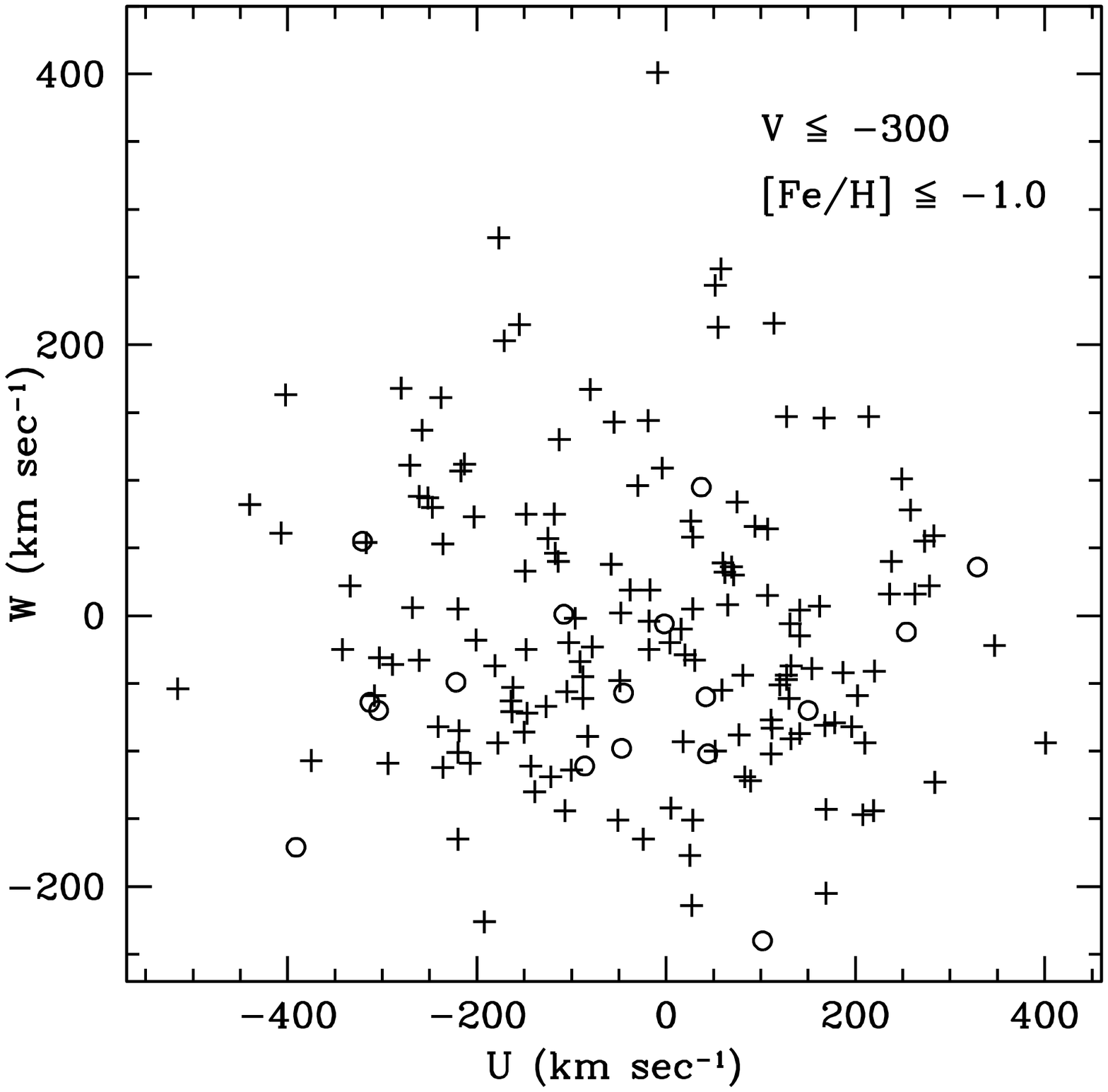}
\caption{The distribution in U and W velocities for single
stars (+) and binary stars (O) with [Fe/H] $\leq\ -1.0$
and V $\leq\ -300$ \kms. \label{fig21}}
\end{figure}
\clearpage

\begin{figure}
\epsscale{1}
\figurenum{22}
\plotone{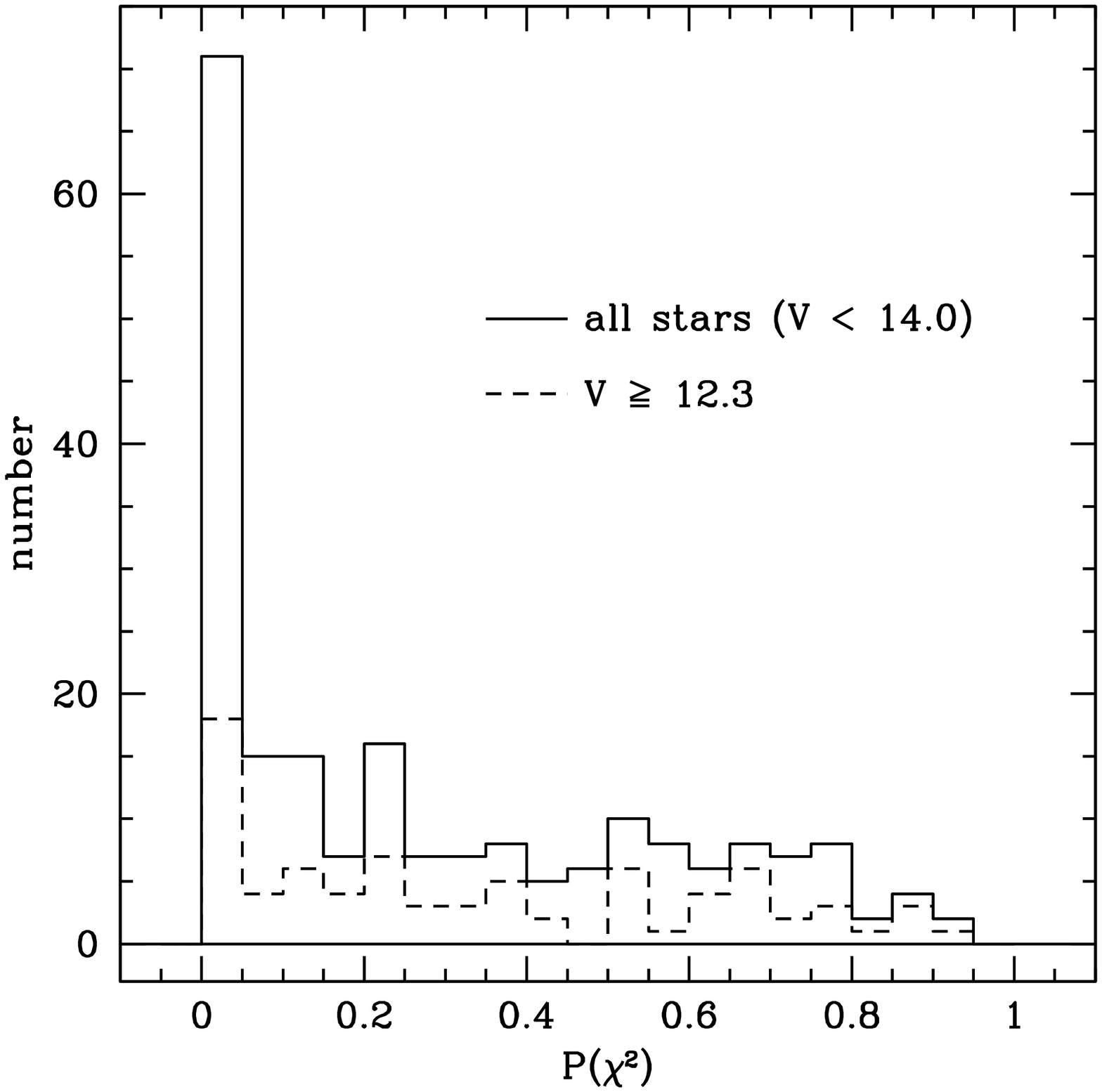}
\caption{The distribution of the P($\chi^{2}$) values
for the probabilities that the variation in measured radial velocities for
red giants in $\omega$~Cen could arise from measurement uncertainties.
The data show evidence for radial velocity variability for
roughly 40\% of the stars, but velocity jitter for the brightest stars
appears to be a major contributing factor. \label{fig22}}
\end{figure}

\clearpage

\begin{figure}
\epsscale{1}
\figurenum{A1}
\plotone{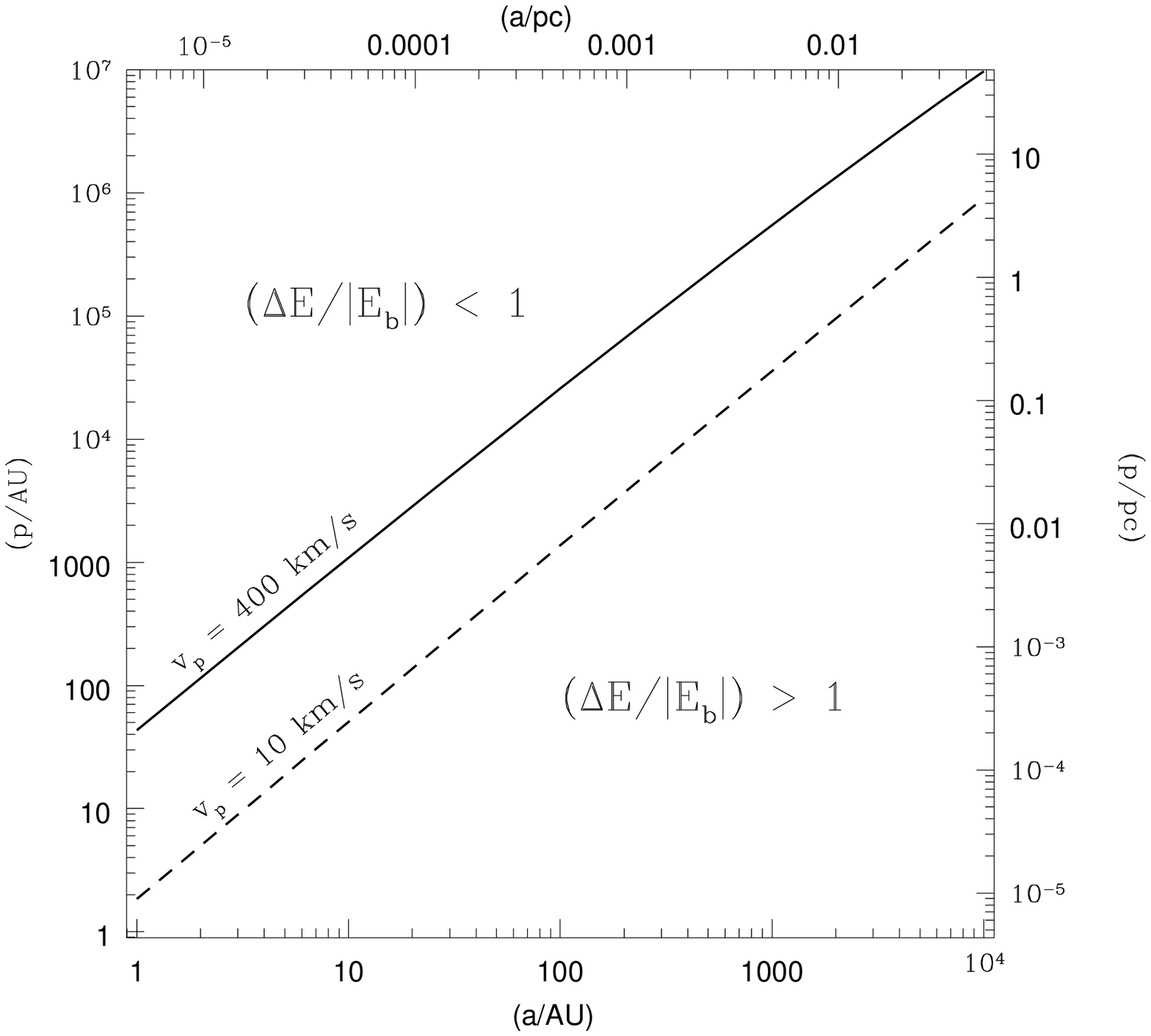}
\caption{Figure A1. Lines corresponding 
to $(\Delta E/|E_b|) = 1$ for encounters with
a $10^6\, M_\odot$ point mass with a relative velocity of 10 or 400 \kms.
Encounters with a larger binary separation $a$ or a smaller impact parameter $b$
are strong encounters that destroy the binary. \label{figa1}}
\end{figure}


\clearpage

\begin{thebibliography}{}

\bibitem[Abt \& Levy (1969)]{abt69} Abt, H.\ A., \& Levy, S.\ G.\
1969, \aj, 74, 908

\bibitem[Abt \& Willmarth (1987)]{abt87} Abt, H.\ A., \&
Willmarth, D.\ W.\ 1987, \apj, 318, 786

\bibitem[Aguilar (1993)]{aguilar93} Aguilar, L.\ 1993, in ASP
Conf.\ Seri.\ 49, Galaxy Evolution:
The Milky Way Perspective, ed. S.\ R.\ Majewski
(San Francisco: ASP), 155.

\bibitem[Aguilar et al.\ (1988)]{aguilar88} Aguilar, L., Hut, P., \&
Ostriker, J.\ P.\ 1988, \apj, 335, 720

\bibitem[Bahcall et al.\ (1992)]{bfg} Bahcall, J.\ N., Flynn, C., 
\& Gould, A.\ 1992, \apj, 389, 234

\bibitem[Burris et al.\ (2000)]{burris2000} Burris, D.\ L., Pilachowski, C.\ A., 
Armandroff, T.\ E., Sneden, C.,
Cowan, J.\ J., \& Roe, H.\ 2000, \apj, 544, 302

\bibitem[Carney et al.\ (1996)]{clla96} Carney, B.\ W., Laird, J.\ B.,
Latham, D.\ W., \& Aguilar, L.\ 1996, \aj, 112, 668

\bibitem[Carney et al.\ (1987)]{cllk87} Carney, B.\ W., Laird, J.\ B.,
Latham, D.\ W., \& Kurucz, R.\ L.\ 1987, \aj, 94, 1066

\bibitem[Carney \& Latham (1986)]{cl} Carney, B.\ W., \&
Latham, D.\ W.\ 1987, \aj, 93, 116

\bibitem[Carney et al.\ (1994)]{paper12} Carney, B.\ W.,
Latham, D.\ W., Laird, J.\ B. \& Aguilar, L.\ A.\ 1994, \aj, 107, 2240

\bibitem[Carney et al.\ (2001)]{bs} Carney, B.\ W.,
Latham, D.\ W., Laird, J.\ B., Grant, C.\ E., \& Morse, J.\ A.\ 2001,
\aj, 122, 3419

\bibitem[Chernoff \& Djorgovski (1989)]{chernokk89} Chernoff, D., \&
Djorgovski, S.\ 1989, \apj, 339, 904

\bibitem[Chernoff et al.\ (1986)]{chernoff86} Chernoff, D., Kochanek, C.,
\& Shapiro, S.\ 1986, \apj, 309, 183

\bibitem[Cohen (2004)]{cohen2004} Cohen, J.\ G.\ 2004, \aj, 127, 1545
 
\bibitem[C\^{o}t\'{e} et al.\ (1996)]{cote96} C\^{o}t\'{e}, P., 
Pryor, C., McClure, R.\ D.,
Fletcher, J.\ M., \& Hesser, J.\ E.\ 1996, \aj, 112,574

\bibitem[Crampton \& Hartwick (1972)]{crampton} Crampton, D.,
\& Hartwick, F.\ D.\ A.\ 1972, \aj, 77, 590

\bibitem[Dinescu et al.\ (1999)]{dana1999} Dinescu, D.\ I.,
Girard, T.\ M., \& van Altena, W.\ F.\ 1999,
\aj, 117, 1792

\bibitem[Duquennoy \& Mayor (1991)]{dm91} Duquennoy, A., \&
Mayor, M.\ 1991, \aap, 248, 485

\bibitem[Eggen (1958a)]{eggen58a} Eggen, O.\ J.\ 1958a, \mnras, 118, 65

\bibitem[Eggen (1958b)]{eggen58b} Eggen, O.\ J.\ 1958b, \mnras, 118, 154

\bibitem[Eggen (1958c)]{eggen58c} Eggen, O.\ J.\ 1958c, \mnras, 118, 560

\bibitem[Eggen (1960a)]{eggen60a} Eggen, O.\ J.\ 1960a, \mnras, 120, 430

\bibitem[Eggen (1960b)]{eggen60b} Eggen, O.\ J.\ 1960b, \mnras, 120, 448

\bibitem[Eggen (1960c)]{eggen60c} Eggen, O.\ J.\ 1960c, \mnras, 120, 540

\bibitem[Eggen (1960d)]{eggen60d} Eggen, O.\ J.\ 1960d, \mnras, 120, 563

\bibitem[Eggen (1977)]{eggen77} Eggen, O.\ J.\ 1977, \apj, 215, 812

\bibitem[Eggen (1959)]{eggen59} Eggen, O.\ J., \& Sandage, A.\ R.\ 1959,
\mnras, 119, 255

\bibitem[Fall \& Zhang (2001)]{fall01} Fall, S.\ M., \& Zhang, Q.\ 2001,
\apj, 561, 751

\bibitem[Freeman (1993)]{freeman93} Freeman, K.\ 1993, in ASP Conf.\ Ser.\ 265,
$\omega$ Centauri: A Unique Window into Astrophysics, ed. F.\ van Leeuwen, J.\ D.\
Hughes, \& G.\ Piotto (San Francisco: ASP), 365

\bibitem[Fulbright (2000)]{fulbright2000} Fulbright, J.\ P.\ 2000, \aj, 120, 1841 (F2000)

\bibitem[Goldberg et al.\ (2002)]{dlsb} Goldberg, D., Mazeh, T., 
Latham, D.\ W., Stefanik, R.\ P.,
Carney, B.\ W., \& Laird, J.\ B.\ 2002, \aj, 124, 1132

\bibitem[Gratton et al.\ (2003)]{gratton03} Gratton, R.\ G., Carretta, E.,
Claudi, R., Lucatello, S., \& Barbieri, M.\ 2003, aap, 404, 187 (M2004)

\bibitem[Greenstein \& Saha (1986)]{greenstein} Greenstein, J.\ L.,
\& Saha, A.\ 1986, \apj, 304, 721

\bibitem[Gunn \& Griffin (1979)]{gg} Gunn, J.\ E.,
\& Griffin, R.\ F.\ 1979, \aj, 84, 752

\bibitem[Harris (1991)]{harris91} Harris, W.\ E.\ 1991, \araa, 29, 543

\bibitem[Helmi et al.\ (1999)]{helmi} Helmi, A., White, 
S.\ D., de Zeeuw, P.\ T., \&
Zhao, H.\ 1999, Nature, 402, 53

\bibitem[Hurley-Keller et al.\ (1999)]{HK99} Hurley-Keller, D.,
Mateo, M., \& Grebel, E.\ K.\ 1999, \apjl, 523, L25

\bibitem[Hut et al.\ (1992)]{hut92} Hut, P., McMillan, S., 
Goodman, J., Mateo, M., Phinney, S., Pryor, C.\ P., Richer, H.\ B.,
\& Weinberg, M.\ 1992, \pasp, 104, 981

\bibitem[Ibata et al.\ (2002)]{ibata02} Ibata, R.\ A., Lewis, G.\ F.,
Irwin, M.\ J., \& Quinn, T.\ 2002, \mnras, 332, 915

\bibitem[Jasniewicz \& Mayor (1986)]{jw86} Jasniewicz, G.,
\& Mayor, M.\ 1986, \aap, 170, 55

\bibitem[Kerr et al.\ (1986)]{kerr86} Kerr, F.\ J., \& Lynden-Bell, D.\
1986, \mnras, 221, 1023

\bibitem[Laird et al.\ (1988a)]{paper3} Laird, J.\ B., Carney, B.\ W.,
\& Latham D.\ W.\ 1988, \aj, 95, 1843

\bibitem[Laird et al.\ (1988b)]{paper7} Laird, J.\ B., Carney, B.\ W.,
Rupen, M.\ P., \& Latham D.\ W.\ 1988, \aj, 96, 1908

\bibitem[Langer et al.\ (1998)]{langer} Langer, G.\ E., Fischer, D.,
Sneden, C., \& Bolte, M.\ 1998, \aj, 115, 685

\bibitem[Latham et al.\ (1988)]{orbs1} Latham, D.\ W., Mazeh, T.,
Carney, B.\ W., McCrosky, R.\ E., Stefanik, R.\ P., \& Davis, R.\ J.\
1988, \aj, 96, 567
 
\bibitem[Latham et al.\ (1992)]{orbs2} Latham, D.\ W., Mazeh, T.,
Stefanik, R.\ P., Davis, R.\ J., Carney, B.\ W., Krymolowski, Y.,
Laird, J.\ B., Torres, G., \& Morse, J.\ A.\ 1992, \aj, 104, 774

\bibitem[Latham et al.\ (1984)]{latham84} Latham, D.\ W.,
Schechter, P., Tonry, J., Bahcall, J.\ N., \& Soneira, R.\ M.\
1984, \apjl, L41
 
\bibitem[Latham et al.\ (2002)]{slsb} Latham, D.\ W., Stefanik, R.\ P., 
Torres, G., Davis, R.\ J.,
Mazeh, T., Carney, B.\ W., Laird, J.\ B., \& Morse, J.\ A.\ 2002,
\aj, 124, 1144

\bibitem[Leon et al.\ (2000)]{leon00} Leon, S., Meylan, G.,
\& Combes, F.\ 2000, \aap, 359, 907

\bibitem[Lindgren et al.\ (1987)]{lindgren} Lindgren, H.,
Ardeberg, A., \& Zuiderwijk 1987, \aap, 188, 39

\bibitem[Lupton et al.\ (1987)]{lupton87} Lupton, R.\ H., 
Gunn, J.\ E., \& Griffin, R.\ F.\ 1987,
\aj, 93, 114

\bibitem[Mayor \& Mermilliod (1997)]{mayormermilliod84} Mayor, M.,
\& Mermilliod, J.-C.\ 1984, in Observational Tests of
Stellar Evolution Theory, IAU Sym.\ No.\ 105, ed.\ A.\ Maeder \&
A.\ Renzini (Reidel, Dordrecht), p.\ 411
 
\bibitem[Mayor et al.\ (1997)]{mayor97} Mayor, M., Meylan, G.,
Udry, S., Duquennoy, A., Andersen, J., Nordstr\"{o}m, B.,
Imebrt, M., Maurice, E., Pre\'{e}vot, Ardeberg, A., \&
Lindgren H.\ 1997, \aj, 114, 1087

\bibitem[Mayor \& Turon (1982)]{mayorturon} Mayor, M.,
\& Turon, C.\ 1982, \aap, 110, 241

\bibitem[McWilliam (1997)]{mcwilliam97} McWilliam, A.\ 1997,
\araa, 35, 503

\bibitem[M\'{e}ndez et al.\ (2000)]{mendez} M\'{e}ndez, R.\ A., 
Platais, I., Girard, T.\ M.,
Kozhurina-Platais, V., \& van Altena, W.\ F.\ 2000, \apj, 524, L39

\bibitem[Meylan \& Heggie (1997)]{meylan97} Meylan, G., \& Heggie, D.\ C.\
1997, \aapr, 8, 1

\bibitem[Meza et al.\ (2004)]{meza04} Meza, A., Navarro, J.\ F.,
Abadi, M.\ G., \& Steinmetz, M.\ 2004, \mnras, in press

\bibitem[Mizutani, Chiba, \& Sakamoto (2003)]{mizutani2003} Mizutani, A.,
Chiba, M., \& Sakamoto, T.\ 2003, \apjl, 589, L89

\bibitem[Norris \& Da~Costa (1995)]{norris95} Norris, J.\ E., \&
Da~Costa, G.\ S.\ 1995, \apj, 447, 680

\bibitem[Odenkirchen et al.\ (2001)]{odenkirchen01} Odenkirchen, M.,
Grebel, E.\ K., Rockoski, C.\ M., Dehnen, W., Ibata, R.,
Rix, H.-W., Stolte, A., Wolf, C., Anderson, J.\ E., Jr., Bahcall,
N.\ A., Brinkman, J., Csabai, I., Hennessy, G., Hindsley, F.\ B.,
Ivezi\'{c}, \v{Z}., Lupton, R.\ B., Munn, J.\ A., Pier, J.\ R.,
Stoughton, C., \& York, D.\ G.\ 2001, \apjl, 548, L165

\bibitem[Odenkirchen et al.\ (2003)]{odenkirchen03} Odenkirchen, M.,
Grebel, E.\ K., Dehnen, W., Rix, H.\-W., Yanny, B., Newberg, H.\ J.,
Rockosi, C.\ M., Marti\'{i}nez, D., Brinkmann, J., \&
Pier, J.\ R.\ 2003, \aj, 126, 2385

\bibitem[Pryor et al.\ (1988)]{pryor88} Pryor, C., 
Latham D.\ W., \& Hazen, M.\ L.\ 1988,
\aj, 96, 123

\bibitem[Rey et al.\ (2004)]{rey2004} Rey, S.\-C., Lee, Y.\-W.,
Ree, C.\ H., Joo, J.\-M., Sohn, Y.\-J., \& Walker, A.\ R.\ 2004,
\aj, 127, 958

\bibitem[Richar \& Fahlman (1986)]{richer86} Richer, H.\ B.\ \&
Fahlmann, G.\ G.\ 1986, \apj, 304, 273

\bibitem[Ryan (1989)]{ryan89} Ryan, S.\ G.\ 1989, \aj, 98, 1693

\bibitem[Ryan \& Norris (1991a)]{ryan91a} Ryan, S.\ G., \& Norris, J.\ E.\ 1991,
\aj, 101, 1835

\bibitem[Ryan \& Norris (1991b)]{ryan91b} Ryan, S.\ G., \& Norris, J.\ E.\ 1991,
\aj, 101, 1865

\bibitem[Siegel et al.\ (2001)]{siegel} Siegel, M.\ H., Majewski,
S.\ R., Cudworth, K.\ M., \& Takamiya, M.\ 2001, \aj, 121, 935

\bibitem[Smith et al.\ (2000)]{smith2000} Smith, V.\ V., Suntzeff, N.\ B,
Cunha, K., Gallino, R., Busso, M., Lambert, D.\ L., \& Straniero, O.\ 2000,
\aj, 119, 1239

\bibitem[Sommer-Larsen (1999)]{sommerlarsen99} Sommer-Larsen, J.\ 1999, \apss,
265, 123

\bibitem[Spitzer (1958)]{spitzer58} Spitzer, L., 1958, \aj, 127, 17

\bibitem[Spitzer (1987)]{spitzer87} Spitzer, L., 1987, Dynamical Evolution
of Globular Clusters (Princeton: Princeton Univ.\ Press)

\bibitem[Stephens \& Boesgaard (2002)]{stephens2002} Stephens, A., \&
Boesgaard, A.\ M.\ 2002, \aj, 123, 1647 (SB2002)

\bibitem[Suntzeff (1993)]{suntzeff93} Suntzeff, N.\ B.\ 1993,
in ASP Conf.\ Ser.\ 48, The Globular Cluster Galaxy Connection, ed.\
G.\ H.\ Smith \& J.\ P.\ Brodie (San Francisco, ASP), 14

\bibitem[Tsuchiya et al.\ (2003)]{tsuchiya2003} Tsuchiya, T., Dinescu, D.\ I.,
\& Korchagin, V.\ I.\ 2003, \apjl, 589, L29

\bibitem[Tsuchiya et al.\ (2004)]{tsuchiya2004} Tsuchiya, T., 
Korchagin, V.\ I., \& Dinescu, D.\ 2004, \mnras, 350, 1141

\bibitem[Vesperini \& Zepf (2003)]{vesperini03} Vesperini, E., \&
Zepf, S.\ E.\ 2003, \apjl, 587, L97

\bibitem[Walker et al.\ (1996)]{walker96} Walker, I.\ R., Mihos,
C., \& Hernquist, L.\ 1996, \apj, 460, 121

\bibitem[Weinberg et al.\ (1987)]{weinberg} Weinberg, M., Shapiro, 
S.\ L., Wasserman, I.\ 1987, \apj, 312, 367

\bibitem[Whitmore (2003)]{whitmore03} Whitmore, B.\ C.\ 2003,
in A Decade of Hubble Space Telescope Science, ed. M.\ Livio, K.\
Noll, \& M.\ Stiavelli (Cambridge: Cambridge Univ.\ Press), in press

\bibitem[Whitmore et al.\ (1999)]{whitmore99} Whitmore, B.\ C.,
Zhang, Q., Leitherer, C., Fall, S.\ M.,
Schweizer, F., \& Miller, B. W.\ 1999, \aj, 118, 1551

\bibitem[Zhang \& Fall (1999)]{zhang99} Zhang, Q., \& Fall, S.\ M.\ 1999,
\apj, 527, L81

\end{thebibliography}
\end{document}